# 3D DNA Origami-Enabled Molecularly Addressable Optical Nanocircuit

*Jaewon Lee*[1,+], *Hayun Ahn*[1,+], *Kyung Hun Rho*[1], *Shelley F. J. Wickham*[2,*], *William M. Shih*[3,4*], and *Seungwoo Lee*[1,3,5,6,*]

[1]KU-KIST Graduate School of Converging Science and Technology, Korea University, Seoul 02841, Republic of Korea

[2]School of Chemistry, School of Physics, The University of Sydney Nano Institute, The University of Sydney, Sydney, NSW 2006, Australia

[3]Wyss Institute for Biologically Inspired Engineering, Harvard University, Boston, Massachusetts 02115, United States

[4]Department of Cancer Biology, Dana Farber Cancer Institute; Department of Biological Chemistry and Molecular Pharmacology, Harvard Medical School, Boston, Massachusetts 02215, United States

[5]Department of Integrative Energy Engineering (College of Engineering) and Department of Biomicrosystem Technology, Korea University, Seoul 02841, Republic of Korea

[6]Center for Opto-Electronic Materials and Devices, Post-Silicon Semiconductor Institute, Korea Institute of Science and Technology (KIST), Seoul 02792, Republic of Korea

*Email: shelley.wickham@sydney.edu.au

*Email: william.shih@wyss.harvard.edu

*Email: seungwoo@korea.ac.kr

+Equally contributed to this work

**Abstract**: Clusters of plasmonic nanoparticles (NPs) offer a versatile platform for controlling electric and magnetic optical resonances, enabling strong light–molecule interactions in applications such as plasmonic resonance energy transfer (PRET). The "optical nanocircuit" concept provides a predictive framework analogous to an electric RLC circuit, where induced dipoles in NPs, ohmic losses in metallic NPs, and dielectric gaps serve as inductors ($L$), capacitors ($C$), and resistors ($R$), respectively. This modular theory allows unprecedented design flexibility, expanding the range of achievable optical resonances in plasmonic clusters. However, existing experimental approaches, such as atomic force microscope tip-enabled nanomanipulation and electron-beam lithography, lack the critical accuracy in nanogap tuning and molecular loading required for applications like PRET. Here, we introduce a molecularly addressable optical nanocircuit enabled by DNA origami. First, we theoretically and experimentally confirmed that gold (Au) NPs and dye-loaded DNA origami can function as different circuit elements: $R$- and $C$-coupled $L$ and $R$-coupled $C$, respectively. To assemble large Au NPs (up to 100 nm) into designer optical nanocircuits, we utilized a mechanically robust 3D DNA origami design rather than conventionally used 2D origami sheet. This platform provided high reproducibility and accuracy in assembling a range of structures—from dimers to tetramers—with controlled symmetry, heterogeneity, and nanogap tunability. Together with ultrasmoothness and uniformity of Au NPs, we achieved the highest $Q$-factor for magnetic resonance of a nanoparticle-based optical nanocircuit (~19.2). Also, selective molecular cargo loading onto designated 3D DNA origami sites within plasmonic clusters enabled deterministic, predictive light-molecule coupling in optical nanocircuits. This resulted in 100-fold stronger PRET signal in dimeric clusters compared to monomeric NPs. Our approach opens promising directions in designing custom optical resonances for use in molecular sensing, nonlinear optics, and quantum photonics.

# 1. Introduction

Localized surface plasmon resonance (LSPR) of noble metal nanoparticles (NPs) has transformed our ability to manipulate light at the deep-subwavelength nanoscale by confining electromagnetic fields below the diffraction limit. This capability enables strong light-matter interactions, driving advances in molecular sensing, nonlinear optics, and quantum photonics (*1–6*). When plasmonic NPs are assembled into clusters—referred to as plasmonic metamolecules—they provide additional control over electric and magnetic resonances (*7–10*). By tuning cluster configuration and interparticle spacing, resonant modes can be precisely tuned; facilitating applications such as plasmonic resonance energy transfer (PRET), Purcell enhancement, which is the accelerated spontaneous emission rate of molecules by its environment, and molecular sensing (*1*, *11–13*). A powerful theoretical framework guiding the design of these plasmonic clusters is the optical nanocircuit model, which establishes an analogy between nanoscale optical components and classical RLC circuit elements (*14*, *15*). RLC electrical circuits consist of inductors (*L*), capacitors (*C*), and resistors (*R*), and play a critical role in filtering and selectively tuning signals at radio frequencies. In the optical model, plasmonic NPs, nanogaps, and material losses (Ohmic losses) can be considered as optical analogs of idealized (i.e., lumped) circuit elements including nanoinductors (*L*), nanocapacitors (*C*), and nanoresistors (*R*), respectively. This approach enables the modular design of plasmonic clusters with precisely tailored electric and magnetic resonances.

The first experimental validation of an optical nanocircuit was demonstrated using dielectric nanorod arrays fabricated by electron-beam lithography (EBL), which functioned as 2D optical circuit elements at mid-infrared wavelengths (*16*). Subsequent experimental approaches have included metallic and dielectric NP clusters assembled via atomic force microscope (AFM) manipulation, demonstrating circuit-like optical behavior at visible wavelengths (*17*). However, these existing methods lack nano-level accuracy in gap size and cluster geometry (e.g., ~75 nm feature size for EBL and 3–8 nm gap size variation for AFM manipulation) and are not scalable. More critically, none of the reported optical nanocircuits to date have demonstrated the molecular addressability required for deterministic coupling between light and specific molecules, or practical applications such as PRET.

In this work, we introduce a DNA origami-enabled optical nanocircuit that is molecularly addressable and scalable. DNA origami is a technique in which a single-stranded DNA 'scaffold' strand is designed to self-assemble into custom 2D and 3D nanoscale shapes by binding of many short DNA 'staple' strands (*18–20*). The staple strands are uniquely addressable and can act as 'handles' on a programmable pegboard for high precision-spatial arrangement of plasmonic NPs and molecules (*21*, *22*). Here we demonstrate that DNA origami NPs can also act as *R*-coupled *C* idealized optical elements. We show that the resistance of this *R* element can be precisely tuned by adjusting the loading of dye molecules on the DNA origami NP. Taking advantage of the molecular addressability of DNA origami, this *R* element can be integrated with plasmonic NP-based *L*- and *C* elements in a highly programmable way. To enable such integration, we designed and used a barrel-shaped 3D DNA origami pegboard. This 3D DNA origami provided increased mechanical robustness, allowing it to accommodate larger and ultrasmooth plasmonic NPs (up to 100 nm) necessary for strong plasmonic resonance. This approach enabled the reliable and accurate assembly of addressable optical nanocircuit architectures, with high precision in NP cluster geometry, interparticle spacing, and molecular loading. As a result, these optical nanocircuits were able to induce unnatural electric or magnetic resonances with the highest *Q*-factor of ~19.2, a value significantly higher than the 4-10 of previously reported plasmonic clusters, matching well with the circuit model-based theoretical predictions: the *Q*-factor, or quality factor, defined as the ratio of power stored to power dissipated in a resonance circuit, reflects the spectral sharpness of resonance–higher *Q* factors correspond to narrower bandwidths and greater spectral selectivity. Importantly, the DNA origami-enabled coupling of dye molecules into this optical nanocircuit allowed for programmable boosting of the light-molecule interactions. For example, the dimeric optical nanocircuits coupled with DNA origami-dye hybrids showed PRET stronger by a factor of 100, as compared to the monomeric counterpart.

## 2. Concept of molecularly addressable optical nanocircuit

The key concept for construction of a molecularly addressable optical nanocircuit are shown in **Figure 1a**. The main components are plasmonic NPs, DNA origami, and dye molecules, which correspond to inductor (*R*-coupled *L*), lossless capacitor (*C* without *R*), and lossy capacitor (*R*-coupled *C*), respectively. As a representative example of plasmonic NPs, we used gold nanospheres (Au NSs), because they can be chemically synthesized into ultrasmooth and uniform plasmonic NPs with high chemical stability that prevents oxidation-induced optical losses (*23–26*). Au NSs can be electrically polarized by light illumination, making them optical *L* elements with simultaneous Ohmic loss (*R* element) during electric dipolar (ED) oscillation (*14*, *15*). Thus, Au NSs can be viewed as *R*-coupled *L* elements. It is important to note that the ratio of resonantly induced scattering to absorption loss of plasmonic NPs, quantified as the ratio of the scattering cross-section (SCS) to the absorption cross-section (ACS), needs to be as large as possible to effectively support complex, higher-order resonance modes. For Au NS, this transition occurs at a diameter of ~ 90 nm. Therefore, we specifically used Au NS larger than this threshold to ensure they could sustain the strong resonances required for our optical nanocircuits. DNA origami components positioned between Au NSs act as an optically transparent dielectric, corresponding to a dielectric capacitor (*C*). Adding dye molecules, which introduce optical loss by absorbing light to the DNA origami, corresponds to adding resistance (*R*), such that a dye-loaded DNA origami can be considered a *R*-coupled *C* element.

In the overall approach here, DNA origami components of optical circuits are used as an addressable molecular pegboard for programming both the location of dye molecules and also the geometry of Au NSs. Modular rearrangement of Au NS, DNA origami, and dye components is predicted to result in a range of different electric/magnetic resonances and PRET responses. Firstly, we designed an optical nanocircuit to independently induce electric (ED, red ball in **Figure 1b**) and magnetic dipoles (MD, blue ball in **Figure 1b**) without coupling between them, which is predicted to have two scattering spectral peaks, or 'bright modes' from ED and MD resonances (**Figure 1c**). Secondly, we designed circuits such that the energy of the induced ED can be partially transferred to induce MD, termed Fano resonance, with a predicted scattering spectral dip in the ED mode at the MD mode wavelength (**Figure 1d**). Finally, we designed circuits to transfer the ED resonance of Au NSs to excite dyes loaded on the DNA origami, which are predicted to result in PRET leading to several scattering spectral dips in the ED mode (**Figure 1e**). Overall, the optical nanocircuit model predicts that on-demand optical resonant modes can be achieved by changing only the combination and spatial arrangement of the core circuit elements.

A barrel-shaped DNA origami structure was used to implement the optical nanocircuit designs (**Figure 1f**). The hollow core of the DNA barrel means that it can be loaded with dye molecules on both inner and outer surfaces, expanding the accessible range for *R*. The barrel's lateral outer surface and top/bottom lids can all be decorated with unique single-stranded DNA

(ssDNA) handles to bind DNA-functionalized Au NSs, allowing for extension along x-, y-, and z-directions.

To achieve optical nanocircuits, the DNA barrels must be structurally rigid enough for the high accuracy-assembly of the relatively large sized Au NSs (up to 100 nm). Even nanometer-scale structural errors can profoundly alter resonant modes in optical circuits (*27, 28*). For example, a few nm error in the nanogap spacing between Au NSs can determine whether Fano resonance is activated or not (*29, 30*). To ensure sufficient mechanical robustness, a three-layered barrel with ~ 10 nm thick walls was used, with an inner diameter of 20 nm, an outer diameter of 30 nm and height of 20 nm (**Figure 1f**; see details in **Figure S1-S2**, Supplementary Information) (*21*). To host more than three 100 nm Au NSs with controlled symmetries on its outer walls, the size ratio between the origami and the Au NPs needed to be ~ 3.3; therefore, a 30 nm origami size was selected. Additionally, the staple strands within the coaxially stacked duplexes were symmetrically arranged to position the Au NS-binding handles on the outer wall of the barrel with two-fold symmetry and uniform vertical alignment (see **Figure S1**, Supplementary Information). This design can also be applied to 3D programmable clustering of Au NPs in future applications.

Using coarse-grained molecular dynamics simulations (oxDNA, **Figure S3**, Supplementary Information), we analyzed the root mean square fluctuation (RMSF) of individual DNA bases within origami barrels (**Figure 1g**). The simulation results, visualized as a colormap overlay on the computed mean structure, reveal that most DNA bases exhibited sub-nm-level RMSF values, confirming the high structural stability of the barrel origami. Notably, RMSF values were approximately 0.5 nm at the center of the sidewall of the DNA origami barrels, where Au NSs are designed to attach. The mechanical stability of the 3-layer DNA barrel is therefore predicted to be sufficient for assembling 100 nm Au NSs for plasmonic clustering with nano-level accuracy.

These results were compared to a two-layered DNA origami barrel with a similar outer diameter, but a thinner wall of ~ 5 nm, as well as 2D DNA origami sheets that have been previously used for plasmonic NP assembly (*31*). The 2-layer barrels and 2D sheets had significantly higher RMSF values > 1 nm (**Figure 1g**), which is expected to lead to nanogap errors and reduced reproducibility in plasmonic nanocircuits. These more flexible DNA origami structures can typically only be used to arrange smaller Au NSs (5~40 nm), which have much weaker plasmonic resonance.

We then theoretically and experimentally confirmed that dye-loaded DNA origami and Au NSs can function as $R$-coupled $C$ and $R$-coupled $L$ circuit elements, respectively. **Figure 2a** shows transmission electron microscope (TEM) images of the assembled barrel DNA origami, with dimensions in good agreement with the design (**Figure S1**, Supplementary Information). More TEM images, further highlighting the scalability and uniformity of the assembled barrel DNA origami, are presented in **Figure S2**, Supplementary Information.

As discussed previously, the dye-loaded barrel DNA origami is designed to function as a lossy nanocapacitor, where the light absorbing properties of dye molecules result in non-radiative loss that is analogous to resistance in an electrical circuit (i.e., $R_{nrad}$, where 'nrad' suffix refers to non-radiative loss, coupled $C$). Simulated electric displacement fields highlight slight converging of an incident electric field into the dye-loaded barrel DNA origami due to excitation of dyes, resulting ED oscillation and emission, as shown in **Figure 2b**. The electric field distribution and power loss density are shown in **Figure S4**, Supplementary Information. The corresponding circuit elements can be represented by a Thevenin equivalent nanocircuit diagram, as shown in **Figure 2c**. Here, the polarization of incident light is assumed to be vertical, so that $R$ and $C$ elements are serially connected under the vertically applied voltage. Each dye-loaded barrel origami can be viewed as an "$R_{nrad}$–$C$" component, where the specific values for these elements are determined from the effective permittivity of the dye-loaded origami (see details in Supplementary Information). YOYO-3 was used as the dye molecule due to its ability to intercalate along DNA duplexes of the barrel through simple mixing and incubation. The experimentally estimated numbers of YOYO-3 dyes intercalated per barrel was ~423 (see details in Supplementary Information).

Next, we synthesized and completed the nanocircuit analyses of Au NSs, as presented in **Figure 2d-2f**. We chemically synthesized ultrasmooth, uniform, and large-sized (100 nm) Au NSs (see TEM images in **Figure 2d**) using iterative growth and etching methods (see details in Supplementary Information) (*23–25*). Low-magnification optical (dark-field) and electron microscope analyses, in conjunction with statistical analyses, was used to characterize the uniformity of Au NSs used in this study (see **Figure S5-S6**, Supplementary Information). We note that in such large sized Au NSs, the resonantly induced scattering, quantified by SCS in ED-based bright mode, can outperform absorptive loss (ACS) (**Figure S7**, Supplementary Information).

The numerically simulated fringe electric displacement field, resonantly induced via ED mode, is shown in **Figure 2e**. Under illumination with vertically polarized light, the divergence and convergence of the newly generated electric fields (white arrows) from top or bottom of Au NSs is denoted the "fringe." Thus, in the corresponding nanocircuit model, $L_P$ coupled with $R$ can be oriented along the light polarization direction (**Figure 2f**). Also, note that $R$ can be divided into two contributions (i) $R_{nrad}$, stemming from Ohmic loss under ED oscillation and (ii) radiative loss ($R_{rad}$), originating from scattering. Thus, each Au NS can be viewed as "$R_{nrad}$– $L_P$ – $R_{rad}$." As EDs, formed at the top and bottom of Au NSs, are coupled to the surrounding air medium, such serially connected "$R_{nrad}$– $L_P$ – $R_{rad}$" elements need to be connected further with a capacitor ($C_f$) in a serial circuit. Note that this $C_f$ corresponds to the fringe in **Figure 2e**.

We then experimentally validated the nanocircuit model predictions of spectral responses for the YOYO-3-loaded origami barrels (**Figure 2g**) and Au NSs (**Figure 2h**). Spectral dispersions of the resistance and scattering were analytically predicted using the nanocircuit model for YOYO-3-loaded barrels and Au NSs (green line **Figure 2g and 2h**, details in Supplementary Information). These analytical predictions were in good agreement with full-3D numerical calculation of ACS and SCS (see blue line in **Figures 2g and 2h**) using finite element method (FEM, see details in Supplementary Information). YOYO-3-loaded origami barrels (**Figure 2g**) were experimentally characterized using UV-Vis absorption spectroscopy, giving two absorption peaks at 613 nm and 575 nm wavelengths (red line **Figure 2g**). The experimental results were in good agreement with both analytical and numerical theoretical predictions.

For Au NSs, the resonant energy exchange between the Au NS and its environment can occur in the form of radiatively generated electric fields, that are also accompanied with the resistance (radiative loss, $R_{rad}$). $R_{rad}$ is serially connected with $L_P$– $R_{nrad}$– $C_f$, as shown in **Figure 2f**. This nanocircuit analytical model predicts the SCS will have induced ED resonance and resultant SCS peak at 570 nm wavelength. This was in good agreement with the numerically predicted FEM results (see green and blue lines of **Figure 2h**). We used dark-field microspectroscopy to experimentally measure the ED-resonant scattering for individual Au NSs (red line of **Figure 2h** and corresponding scattering image in inset of **Figure 2h**, see Supplementary Information). The measured scattering spectrum matched well with both nanocircuit- and FEM-based theoretical prediction. The nanocircuit parameters used in **Figure 2** are summarized in **Table S1**. Overall, the two core components of our nanocircuit, the dye-loaded origami barrels and Au NSs, were confirmed to act as first-order optical nanocircuits.

## 3. Programmable self-assembly of higher-order optical nanocircuit

Following characterization of the individual components, we next increase the complexity of the optical nanocircuit using the molecular addressability of the outer walls of DNA origami barrel. We assembled dimeric nanocircuits (second order) and experimentally quantified their resonant optical properties, as summarized in **Figure 3**. The outer origami barrel can be symmetrically divided into four quadrants (**Figure 3a**), each of which can be programmed with a unique ssDNA handle sequence. The molecular addressability of each ssDNA handle was confirmed by super-resolution optical imaging (DNA-PAINT), as shown in **Figure 3b** (**Figure S8**, details in Supplementary Information). The 100 nm Au NSs were functionalized with the complementary 'anti-handle' ssDNA via thiol-Au chemistry (see Supplementary Information).

First, we assembled two Au NSs into a symmetric dimer configuration with the barrel origami positioned at the center, as shown in **Figure 3c**. The gap between Au NSs is predicted to be diameter of barrel origami, which is 30 nm. Using gel-electrophoresis, symmetric dimers were purified from free Au NSs and barrel (**Figure 3d**). Monodispersed dimers, composed of two 100 nm Au NSs, were obtained with a high yield, as observed by TEM (**Figure 3e**). Theoretical spatial distributions of electric displacement field distributions at ED modes under horizontal (**Figure 3f**) and vertical (**Figure 3g**) illuminations were then numerically simulated.

Horizontal illumination can give rise to the capacitive coupling between two Au NSs, resulting in an ED mode at 700 nm wavelength and an electric quadrupolar mode (EQ) at 550 nm wavelength. Thus, the equivalent nanocircuit model (**Figure 3h**) has to include both $C_f$, responsible for the fringe of ED, and also $C_{ab}$, responsible for capacitive light confinement at center-positioned dielectric barrel origami, denoted as hot spots. As the ED is formed along the direction of light polarization, these $C_f$ and $C_{ab}$ need to be arranged in a series circuit. Also, $C_{ab}$ is located between the two Au NSs (inductors) to reflect the capacitive coupling. As mentioned above, each Au NS, highlighted by "A" and "B" (yellow box **Figure 3h**), acts as "$R_{nrad} - L_P - R_{rad}$." The theoretical predictions were validated by the polarization-dependent, dark-field microspectroscopy scattering spectrum measurement of an individual dimer (**Figure 3i**, red line), whose specific geometry and orientation were predetermined using correlative TEM imaging (details in **Figure S9**, Supplementary Information).

Note that the nanocircuit model only captures the ED resonance, not the higher-order EQ peak. This is a fundamental limitation, as the lumped circuit elements for each NPs are derived from a dipolar quasistatic approximation. Therefore, the model cannot describe the EQ mode in this dimer, that involve non-dipolar charge oscillations within the individual NPs (see **Figure S10** in Supplementary Information). The full-wave FEM simulation correctly predicts the EQ peak, although its position appears at a slightly shorter wavelength than the measured value, likely due to minor geometric differences in the model compared to the assembled structures. The $Q$-factor of ED scattering was found to be approximately 5.9 (details in **Figure S11**, Supplementary Information).

In contrast, vertically polarized illumination enables independent ED resonances in each Au NSs without coupling. In experimental measurements of the scattering spectrum, this resulted in excitation of only ED at the shorter wavelength of 565 nm (red line **Figure 3k**). The corresponding $Q$-factor was found to be approximately 3.51 (details in **Figure S11**,

Supplementary Information). For vertical illumination, "$R_{nrad} - L_P - R_{rad}$." element of Au NS and $C_{ab}$ can be arranged in a parallel circuit (**Figure 3j**). Similarly to the horizontal case, the vertical illumination nanocircuit model and FEM simulation both agreed well with the dark-field scattering measurement (**Figure 3k**).

In addition to symmetric dimers, asymmetric dimers were also assembled. In this variant, a 40 nm Au NS was assembled at an adjacent quadrant of the barrel origami to a 100 nm Au NS, as shown in **Figure 4a**. Based on the designed 90° spacing of barrel handles between quadrants, the gap between the 40 nm and 100 nm Au NSs is expected to be ~5 nm in this configuration. The spatial distributions of electric displacement field distributions at ED modes under horizontal (**Figure 4b**) and vertical (**Figure 4c**) illumination were numerically simulated. These distributions exhibit a general trend similar to that found in symmetric dimers. Thus, the nanocircuit diagrams for vertically and horizontally polarized illuminations were the same as those for symmetric dimers (**Figure 3h** and **3j**). The model predictions agreed well with the FEM simulations and dark-field scattering spectra (**Figure 4d and 4e**). Similar to the case of the symmetric dimer, however, the dipolar circuit model does not capture the higher-order EQ mode (see **Figure S10** in Supplementary Information). For the asymmetric system the ED resonances occur at shorter wavelengths than the symmetric system (for horizontal polarization: from 700 nm to 633 nm wavelength; for vertical polarization from 565 nm to 552 nm wavelength), due to the smaller size of one of the Au NSs. The asymmetric dimer had a higher measured $Q$-factor of ED scattering compared with the symmetric dimer (9.75 compared to 3.51). This increase in $Q$-factor of ED scattering could be due to both decreased radiative loss from the smaller Au NS and the reduced gap between the Au NSs (details in **Figure S11**, Supplementary Information). Indeed, $R_{rad}$ of the asymmetric dimer was found to be smaller than for the symmetric dimer (see details on nanocircuit parameters, summarized in **Table S2**, Supplementary Information).

Third and fourth order nanocircuits consisting of trimeric and tetrameric assemblies were then assembled. A combinatorial approach using 40 nm and 100 nm Au NSs, allowed for assembly of trimers and tetramers with controlled symmetry and heterogeneity (**Figure 5a-5d**). For instance, (i) two 100 nm Au NSs and one 40 nm Au NS with mirror symmetry and heterogeneity (**Figure 5a**); (ii) two 40 nm Au NS and one 100 nm Au NS with mirror symmetry and heterogeneity (**Figure 5b**); (iii) three 100 nm Au NSs with mirror symmetry without heterogeneity (**Figure 5c**); (iv) two 100 nm Au NSs and two 40 nm Au NSs with mirror symmetry (**Figure 5d**). The assembled and gel-purified nanocircuits showed a high reproducibility, with low variability in the measured nanogap between Au NSs (see statistical analyses in **Figure 5e**).

Higher order nanocircuits enable access to two important optical modes that would otherwise be impossible to achieve. Firstly, clusters of more than 2 NPs can complete a ring of three capacitors, which can induce unnatural magnetic resonance via capacitively induced circulation of the electric field displacement (**Figure 6a**). As representative examples, we include the MD mode analyses for cases (iii) and (iv) in **Figure 6a,** showing the circulating electric displacement field. Regardless of the polarization of illumination, ring-like configurations of Au NS inductors ("$R_{nrad} - L_P - R_{rad}$" elements, labeled A, B, and C for trimers; A, B, C, and D for tetramers) and capacitors ($C_{ab}$, $C_{ac}$, and $C_{bc}$, for trimers; $C_{ac}$, $C_{ad}$, $C_{bc}$, and $C_{bd}$, for tetramers) occur in both trimeric and tetrameric nanocircuits (**Figure 6b**).

Experimental measurements of scattering spectra for ring resonators showed strong MD-resonant modes in visible wavelengths, in good agreement with theoretical predictions

(**Figures 6c-f**, see also **Table S3-S4**, Supplementary Information for nanocircuit parameters). In the cases of (i), (ii), and (iv), scattering shoulders in addition to the ED-driven scattering peaks (**Figures 6c, 6d, and 6f**) serves as evidence that the MD resonance is induced independently without coupling to ED. We benchmarked these results against relevant literature results (**Figure 6g**) and found that our DNA barrel Au NS metamolecule trimer (i) achieved the highest MD mode $Q$-factor to date (**Figure S12**, Supplementary Information) (*8, 9, 32, 33*).

Secondly, metamolecule (iii) demonstrated Fano-resonance, where the energy of the 'bright' ED mode is partially transferred to resonantly induce a 'dark' MD mode (details in **Figure S13**, Supplementary Information). The mirror symmetry of this trimeric nanocircuit is broken when the incident polarization is horizontal because the gaps between 1–2 and 2–3 Au NSs is about 1 nm, while 1 and 3 Au NSs are separated by ~20 nm. This resulted in Fano-resonance with record high $Q$-factor compared to previous work (**Figure 6e and 6h**) (*24, 29, 31*). These Fano resonances were also well predicted with circuit models with the appropriate parameters (summarized in **Table S3**, Supplementary Information). We attribute our improved $Q$-factors to increased stability and accuracy of the DNA barrel and the smoothness of the Au NSs in our optical nanocircuits. Previously, self-assembled plasmonic Fano resonators have been based on closely packed tetramers or heptamers where the higher number of NPs per cluster leads to greater radiative loss and therefore decreases the $Q$-factor.

## 4. Digitally controlled and boosted PRET using molecular addressability

Lastly, we increased the efficiency of PRET with our molecularly addressable optical nanocircuits. In particular, we contrasted PRET efficiencies between monomeric and dimeric optical nanocircuits assembled with 100 nm Au NSs (**Figure 7a**), revealing a large difference in electric field confinement and light-molecule coupling (**Figure 7b and 7c**). To achieve this, we experimentally and theoretically quantified the threshold molecular loading of YOYO-3 (i.e., $R_{nrad}$), beyond which PRET can be induced. To precisely control the amount of YOYO-3 loading, we immobilized the Au NS monomers and assembled symmetric Au NS dimers in a microfluidic chamber and adjusted the concentration of YOYO-3 dye molecules (see Supplementary Information). The Au NS monomer was conjugated with double stranded DNA (dsDNA) to facilitate the loading of YOYO-3 on the Au NS via dye molecule's strong affinity to intercalate with dsDNA. In contrast, the surface of Au NS for dimeric circuits were modified with only single stranded DNA (ssDNA), such that the YOYO-3 dye would preferentially load on the dsDNA of the origami barrel for this sample.

The Thevenin equivalent nanocircuit diagrams for dye-loaded monomeric and dimeric circuits are shown in **Figure 7d and 7e**, respectively. Note that the sub-circuit, where $R_{nrad}$ is connected with $C_c$ and $C_f$ or $C_{AB}$, needs to be additionally connected in series with $L$. The $R_{nrad}$ and $C_c$ values were determined by modeling the dsDNA shell of the Au NS monomer and the barrel of the Au NS dimer as effective spheres with varying amounts of YOYO-3 loading (see Supplementary Information). According to the nanocircuit model (**Figure 7d and 7e**), the dye molecules introduce a non-radiative loss channel into the plasmonic resonator. This manifests as a distinct dip in the scattering spectra. We therefore define the PRET-threshold as the minimum number required to produce a spectral dip that is reliably detectable above our measurement noise (corresponding to a PRET efficiency ~ 0.04). For dimers, this threshold was predicted to be ~60 molecules, which was much lower than that the ~2500 predicted for monomers (**Figure 7f**) (circuit parameters summarized in **Table S5**, Supplementary Information). This theoretical prediction was confirmed using dark-field microspectroscopy

(**Figure 7g-k**). Here, a clear PRET dip was observed with ~ 70 YOYO-3 molecules for the dimeric nanocircuit, whereas the monomeric counterpart required 100-fold greater number of molecules (~ 6400) to produce comparable spectral dip, in agreement with both FEM- and nanocircuit model-based theoretical predictions.

## 5. Conclusion

Taken together, our study has achieved three key advances. First, we demonstrated that DNA origami and Au NSs can each be interpreted as idealized lumped optical circuit elements, enabling the programmable and modular design of plasmonic resonators. This conceptually bridges molecular nanofabrication with circuit-based photonic design, offering a promising framework for building functional optical nanostructures with molecular precision.

Second, by utilizing mechanically robust 3D DNA origami barrels, we enabled the high-accuracy, high-fidelity self-assembly of optical nanocircuits, capable of supporting high-$Q$ resonances, including both bright modes (ED and MD resonances) and dark modes (MD-based Fano resonance). For example, the mechanical stability of the DNA barrel made it possible to successfully self-assemble non-close-packed, asymmetric trimers, which had previously been challenging to achieve using DNA origami platforms. This greater structural stability also allowed us to successfully assemble large (up to 100 nm) Au NSs into higher order plasmonic clusters with increasing complexity, while maintaining high accuracy and fidelity. In addition, the ultrasmooth and highly spherical features of the Au NSs further improved accuracy of the assembly, when compared to randomly faceted Au NSs prevalently used in DNA origami enabled assembly of plasmonic clusters thus far (*31*, *32*, *34*, *35*). As such, we were able to achieve both bright and dark (Fano) modes with higher $Q$-factors at shorter visible wavelengths than previously reported.

Third, the unique molecular addressability enabled by DNA origami significantly enhanced the interaction between dye molecules loaded onto DNA origami and plasmonic clusters, greatly increasing PRET efficiency. In the dimeric optical nanocircuit, PRET was induced with a few tens of dye molecules with over 100 times higher efficiency compared to the monomeric counterpart. This demonstrates the feasibility of single-molecule level detection, which has been previously unattainable in conventional PRET-based molecular sensing (*11*, *12*, *36*).

While this work highlights several representative motifs (**Figure 3-7**) of higher-order optical nanocircuits, our approach is broadly extensible. By varying the size of the Au NSs and reprogramming the arrangement of ssDNA handles on the DNA origami barrel, a vast array of nanocircuit geometries can be constructed, enabling continuous increases in structural and functional complexity.

Thus, many previously inaccessible optical resonances can come to the fore through our DNA origami-based strategy. This precise and modular control over optical nanocircuits at the molecular level opens up new opportunities across multiple application domains–not only in single-molecule-level spectroscopy such as PRET, but far beyond. For instance, the ability to fine-tune resonance characteristics such as $Q$-factor and the spatial distributions of electric and magnetic fields enables highly efficient light confinement at the nanoscale–or even in pico-volume domains. This unprecedented level of light focusing allows for light-matter interactions with atoms and molecules at degrees of freedom that were previously unattainable using conventional nanofabrications. As such, our platform has the potential to reshape nonlinear and quantum plasmonics. Furthermore, when combined with emerging DNA origami placement technologies, our approach offers a powerful route for integrating quantum emitters with engineered plasmonic modes at precisely defined spatial locations on chip. This capability paves the way for constructing hybrid quantum systems and scalable quantum optical circuits with molecular level precision.

**Methods**

**Folding DNA origami:** Barrel-shaped DNA origami structures were folded using a p7249 scaffold ssDNA strand (Guild Bioscience Inc.) and a set of complementary staple ssDNA strands (Bioneer) (*21*, *37*). The folding mixture was prepared with 20 nM scaffold, a 10-fold excess of staple strands (200 nM), 16 mM $MgCl_2$, 5 mM Tris-HCl (pH 8.0), and 1 mM EDTA. This mixture was thermally annealed in a Tetrad 2 Peltier thermal cycler (Bio-Rad) using the following protocol: an initial step at 80°C for 15 minutes, followed by a slow cooling ramp from 60°C to 24°C over 36 hours, and a final hold at 4°C. After annealing, the folded DNA origami structures were purified by agarose gel electrophoresis. Samples were mixed with a 6X agarose gel loading buffer (Bioneer) and loaded onto a 1.5 wt% agarose gel prepared with 0.5X Tris-Borate-EDTA (TBE) buffer (Sigma-Aldrich) and 10 mM $MgCl_2$. Electrophoresis was performed in a horizontal system (Bio-Rad) at 60 V for 3 hours. The desired bands, visualized with SYBR Gold Nucleic Acid Gel Stain, were excised from the gel. Finally, the purified origami structures were extracted from the gel slices using Freeze 'N Squeeze™ DNA Gel Extraction Spin Columns (Bio-Rad).

**Synthesis of Au NSs:** Au NSs with diameters of 40 nm and 100 nm were synthesized using a multi-step method based on iterative reductive growth and oxidative dissolution (*23–25*). The process began with the preparation of an Au seed solution by reducing Gold (III) chloride trihydrate ($HAuCl_4$) with ice-cold sodium borohydride ($NaBH_4$). These seeds were then used to grow single-crystal Au nanorods (NRs) in the presence of silver nitrate ($AgNO_3$) as a catalyst and L-ascorbic acid as a reducing agent. The core of the method involves an iterative growth and dissolution process. The Au NRs were first subjected to an oxidative etching step with a controlled concentration of $HAuCl_4$ to form intermediate seeds. These seeds were subsequently used for a further growth step with L-ascorbic acid to form concave rhombic dodecahedra (CRDs). A final etching step, again with $HAuCl_4$, transformed the CRDs into highly uniform and spherical 40 nm Au NSs. To achieve the larger 100 nm size, this growth-and-etch cycle was repeated using the 40 nm Au NSs as seeds. The complete, detailed protocol is available in the Supplementary Information.

**Preparation of DNA functionalized Au NSs:** The surfaces of the Au NSs were functionalized with thiolated anti-handle ssDNA, which is complementary to the handle strands on the DNA origami scaffold. After washing the as-synthesized Au NSs to remove residual surfactant, they were incubated with the thiolated DNA. To maximize the DNA loading density on the nanoparticle surface, a gradual salt-aging process was employed over two days, during which the salt concentration was slowly increased to 750 mM to screen electrostatic repulsion (*25*, *38*, *39*). Successful conjugation was confirmed by observing a characteristic redshift in the plasmon resonance peak of the Au NSs. Finally, the DNA-functionalized Au NSs were washed via centrifugation to remove excess, unbound DNA strands before use. The full, detailed protocol is available in the Supplementary Information.

**Assembly of Au NSs cluster and purification:** To assemble the Au NS clusters, the washed DNA-functionalized Au NSs (DNA-Au NSs) were first redispersed in a buffer containing 10 mM $MgCl_2$, 0.05 wt% SDS, and 10 mM Tris-HCl. These DNA-Au NSs were then mixed with purified barrel DNA origami. The stoichiometric ratio of DNA-Au NSs to barrel DNA origami was set to 10:1, 15:1, and 20:1 for the assembly of dimers, trimers, and tetramers, respectively. These ratios were kept consistent regardless of the Au NS diameter. The concentration of the barrel DNA origami was fixed at 0.5 nM for all experiments. The resulting mixture was incubated overnight at room temperature in a mixer to facilitate cluster formation.

Following incubation, the assembled Au NS clusters were analyzed and purified by agarose gel electrophoresis, as shown in Figures 3 and 5 of the main manuscript. An 0.85% agarose gel was prepared with a buffer containing 0.5X TBE, 10 mM $MgCl_2$, and 1x SYBR Gold Nucleic Acid Gel Stain. Before loading, samples were mixed with a 6x gel loading buffer containing 40% sucrose, 0.05% bromophenol blue, and 0.05% xylene cyanol FF. Electrophoresis was performed at 60 V for 3 hours. The separated bands were visualized with a white LED lamp and imaged using a Gel Doc XR+ gel documentation system (Bio-Rad). The bands corresponding to the well-assembled Au NS clusters were then physically excised from the gel with a razor blade. Finally, the clusters were recovered from the gel slice by squeezing it between two glass slides covered with clean Parafilm.

**Dark field scattering spectra measurement:** The scattering spectra of individual nanoclusters were measured using a custom-built dark-field microspectroscopy setup. To enable direct correlation between a geometry of clusters and its optical response, samples were placed on a Formvar-coated TEM grid, which allowed for TEM imaging of a specific cluster before its spectrum was measured. To investigate the resonant scattering behavior of the Au NS assemblies upon differently polarized light illumination, polarization-resolved spectra were acquired by rotating a linear polarizer in the illumination and detection path relative to the cluster's main axis. For the PRET experiments, a microfluidic chamber was used to immobilize the nanostructures on a glass surface, which permitted the in-situ exchange of solutions with varying dye concentrations during spectral acquisition. The full details of the optical setup and sample preparation are provided in the Supplementary Information.

**Estimation of YOYO-3 loading:** The number of YOYO-3 dye molecules loaded onto the DNA nanostructures was estimated using the McGhee-von Hippel model (*40*). To use this model accurately, we first experimentally determined the association constant of dye molecule ($K_a$) for our specific buffer conditions. This was achieved by measuring the absorption spectra of solutions containing a fixed concentration of dsDNA that were titrated with varying amounts of YOYO-3. The concentration of bound dye at each step was quantified by deconvoluting the distinct spectral contributions of the free and bound dye populations. The resulting $K_a$ value was then used to precisely calculate the dye loading for the main PRET experiments. The full details of this procedure are provided in the Supplementary Information.

**FEM based electromagnetic numerical calculation:** The optical properties of the Au NS clusters were simulated using the finite element method (FEM) in a commercial software package (CST Microwave Studio). The geometry of each simulated cluster, including nanoparticle sizes and inter-particle gap distances, was based directly on experimental TEM images. The clusters were illuminated with a polarization-controlled plane wave at a 53° angle of incidence. The dielectric function of Au was taken from literature values (*41*), while the dielectric properties of the DNA origami and the ssDNA shells on the nanoparticles were calculated based on their composition. For simulations of the PRET system, the dye-loaded DNA components were modeled using a Lorentzian oscillator function whose parameters were derived from experimental absorption spectra. The full details of all simulation parameters, material models, and governing equations are provided in the Supplementary Information.

**Analytical analysis of optical nanocircuit model:** To analyze the scattering and resonant behavior of the Au NS clusters and the PRET system, an optical nanocircuit model was used based on established frameworks (*14*, *17*). Following the simulated electric displacement fields, individual Au NSs were modeled as an inductor (*L*) with radiative and non-radiative resistors (*R*), while the dye-loaded DNA origami was modeled as a lossy capacitor ($R_{nrad}$-*C* component). These components were integrated into a larger circuit to model a full cluster, with additional capacitors and mutual inductors accounting for the near-field coupling between nanoparticles. The final scattering response of the system was then estimated by calculating the total current flowing through the equivalent circuit. The specific details for parameter determination and all circuit parameter values are summarized in the Supplementary Information (**Tables S1-S5**).

**DNA-PAINT analysis:** For DNA-PAINT imaging, a custom-built total internal reflection fluorescence (TIRF) microscope was used. For imaging, the barrel DNA origami was functionalized with docking strands for super-resolution imaging and with biotin molecules for surface attachment. The samples were immobilized within a fluidic chamber via a standard biotin-streptavidin linkage. Imaging was performed by introducing a solution containing 5 nM of fluorescently-labeled imager strands, which are complementary to the docking strands on the origami. The transient binding of these imager strands creates the blinking signal required for super-resolution imaging. Image sequences of 10,000 frames were recorded and then processed using the Picasso software package to reconstruct the final images from the accumulated single-molecule localizations (*42*). The full details of the optical setup and imaging parameters are provided in the Supplementary Information.

**Data availability:** The data generated during this work are available from the corresponding author on reasonable request.

**Author contributions:** S.L. and W.M.S. conceived the original research idea at the initial stage and designed the work. J.L. theoretically analyzed and experimentally realized the optical nanocircuits. H.A. assembled the optical nanocircuits. J.L. and H.A. characterized the spectral responses of the assembled optical nanocircuits. K.H.R. contributed to structural analyses of the assembled optical nanocircuits. W.M.S., and S.F.J.W. designed and experimentally validated barrel-shaped DNA origami. S.L. supervised the research. J.L., W.M.S., S.F.J.W., and S.L. contributed to the writing and editing of the manuscript.

**Acknowledgement**: J.L, H.A, K. H. R., W. M. S, and S. L. acknowledge funding from Korea-US Collaborative Research Fund (KUCRF), funded by the Ministry of Science and ICT and Ministry of Health & Welfare, Republic of Korea (grant number: RS-2024-00468463). J.L, H.A, K. H. R., and S. L. acknowledge funding from National Research Foundation of Korea (NRF-RS-2023-00272363), and Korea University Grant. S. F. J. W. acknowledge funding from the Australian Research Council (ARC DE180101635).

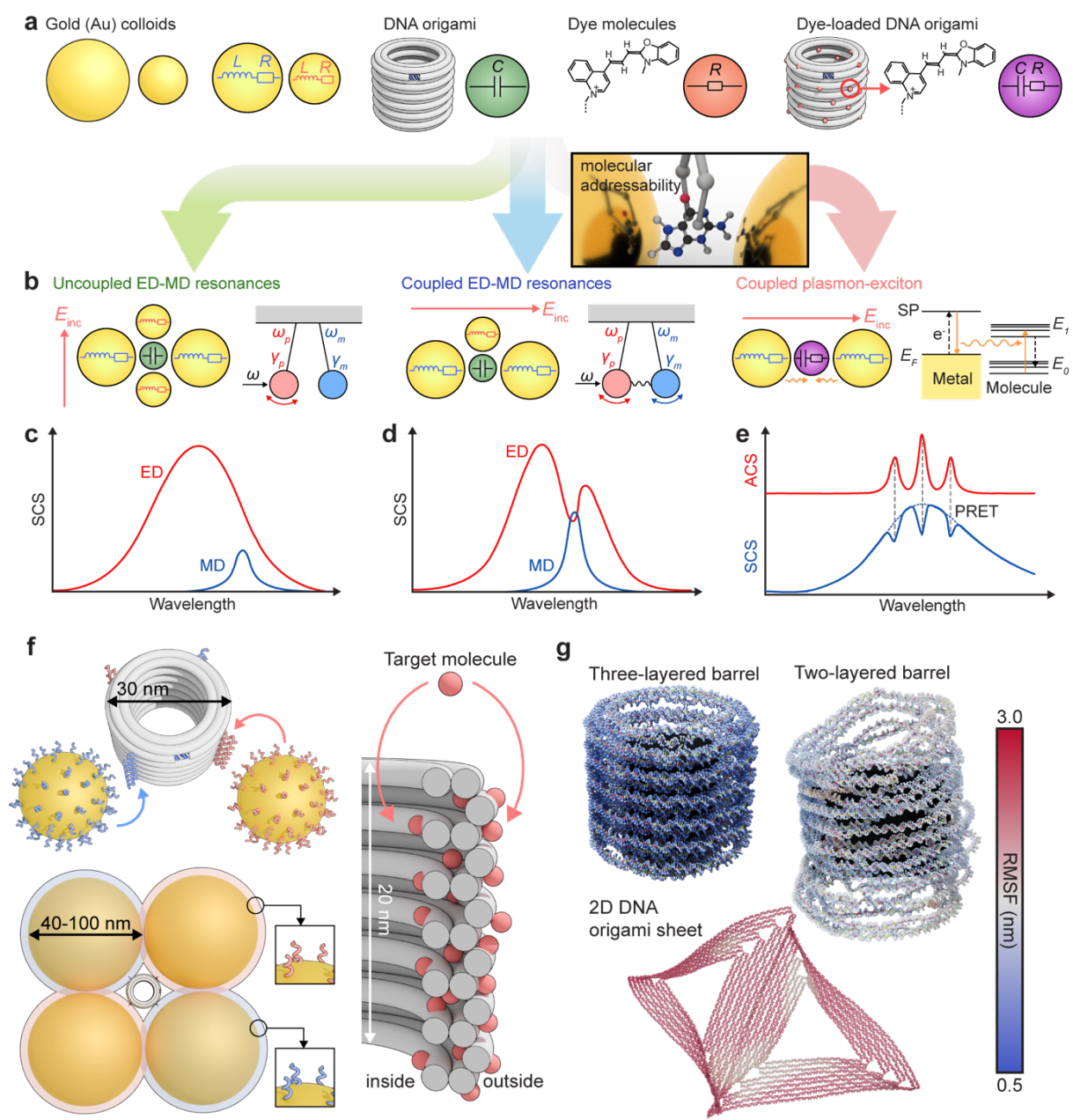

**Fig. 1 | Construction of the molecularly addressable optical nanocircuit. a,** Schematics illustrating the molecularly addressable integration of optical nanocircuit elements (e.g., gold (Au) colloids, DNA origami, dye molecules, dye-loaded DNA origami) for controlling the complex interaction between the optical components. *L*, *R*, and *C* indicate inductor, resistor, and capacitor, respectively. **b,** Representative examples of circuit element arrangement with on-demand plasmon resonance modes. **c-e,** Corresponding scattering response for (**c**) independently induced electric dipole (ED) and magnetic dipole (MD) modes, (**d**) interfering ED and MD mode, and (**e**) energy transfer between ED of Au colloids and molecular dipole (plasmonic resonance energy transfer, PRET). **f,** Schematic illustration of the molecular addressability of the barrel-shaped DNA origami structure on its inner and outer surfaces. Single-stranded DNA (ssDNA) handles (red and blue helices) decorated on the surface act as molecularly programmable docking points for hosting target materials, such as Au NPs labeled with complementary ssDNA. **g,** Simulated mean structure and root mean square fluctuation

(RMSF) of the three-layered barrel DNA origami used in this work, compared to a thin-walled counterpart (two-layered barrel) and to 2D DNA origami sheets.

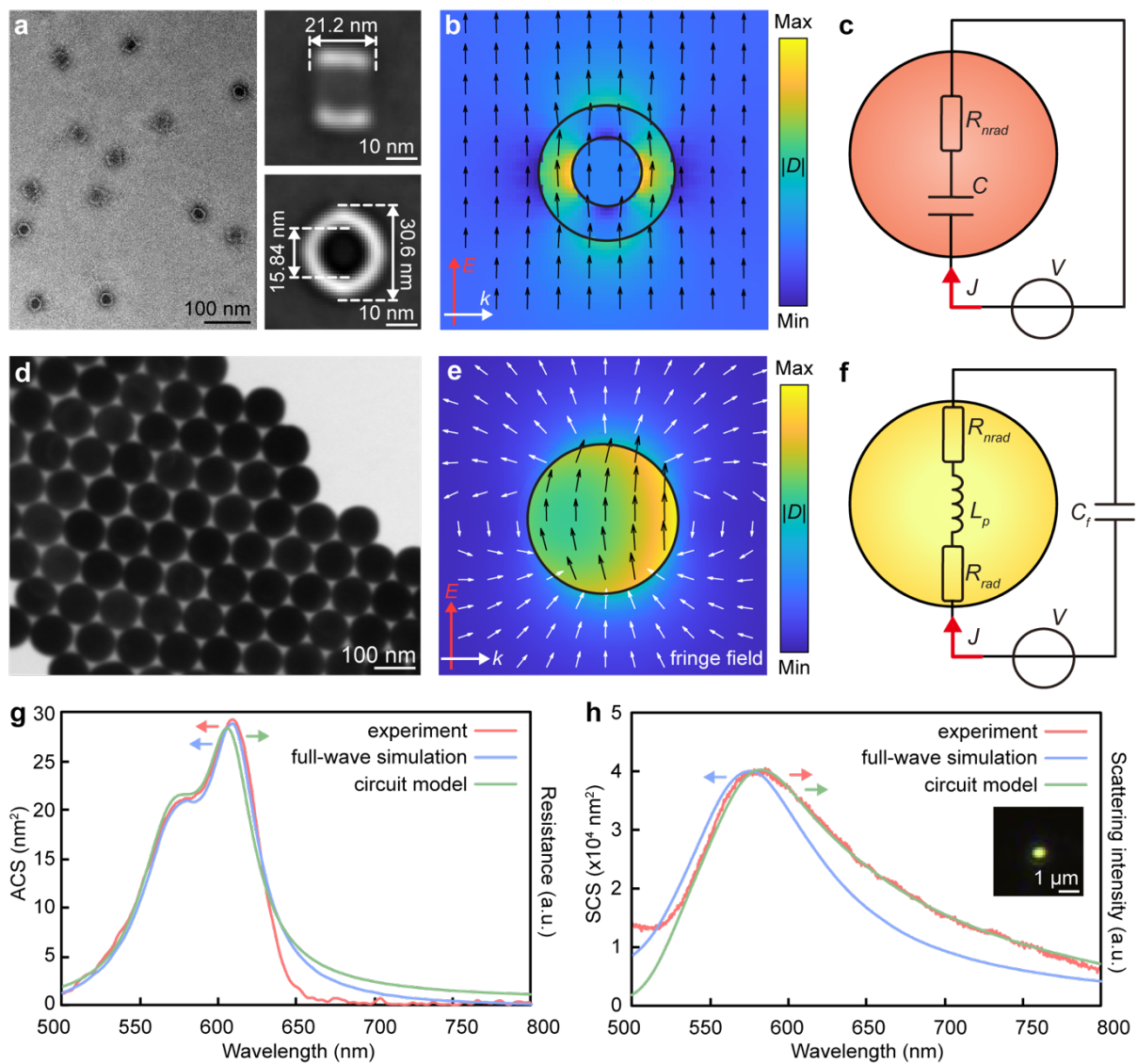

**Fig. 2 | Basic circuit elements for optical nanocircuitry. a,** (left panel) Negative-stained wide-view and (right panel) particle averaged transmission electron microscope (TEM) images of barrel-shaped DNA origami. **b,** Electric displacement field distribution of dye-loaded DNA origami at the wavelength of 610 nm. **c,** Thevenin equivalent nanocircuit diagram for dye-loaded DNA origami. $V$ and $J$ indicate constant voltage source and current flow in the circuit, respectively. **d,** Representative TEM image of Au nanospheres (NSs). **e,** Electric displacement field distribution of a 100 nm Au NS at its ED mode. The white arrows represent fringe field, created by the dipolar resonance of Au NS. **f,** Thevenin equivalent nanocircuit diagram for an Au NS. **g,** Absorbance spectra of YOYO-3 dye-loaded DNA origami obtained via experimental measurement (red line), full-wave simulation (absorption cross-section, ACS, blue line), and analytical prediction using the nanocircuit model (green line). **h,** Scattering spectra of a 95 nm single Au NP acquired via dark-field (DF) scattering (red line), full-wave simulation (scattering cross-section, SCS, blue line), and analytical prediction using nanocircuit model (green line). The inset shows a DF optical microscope image of the characterized Au NS.

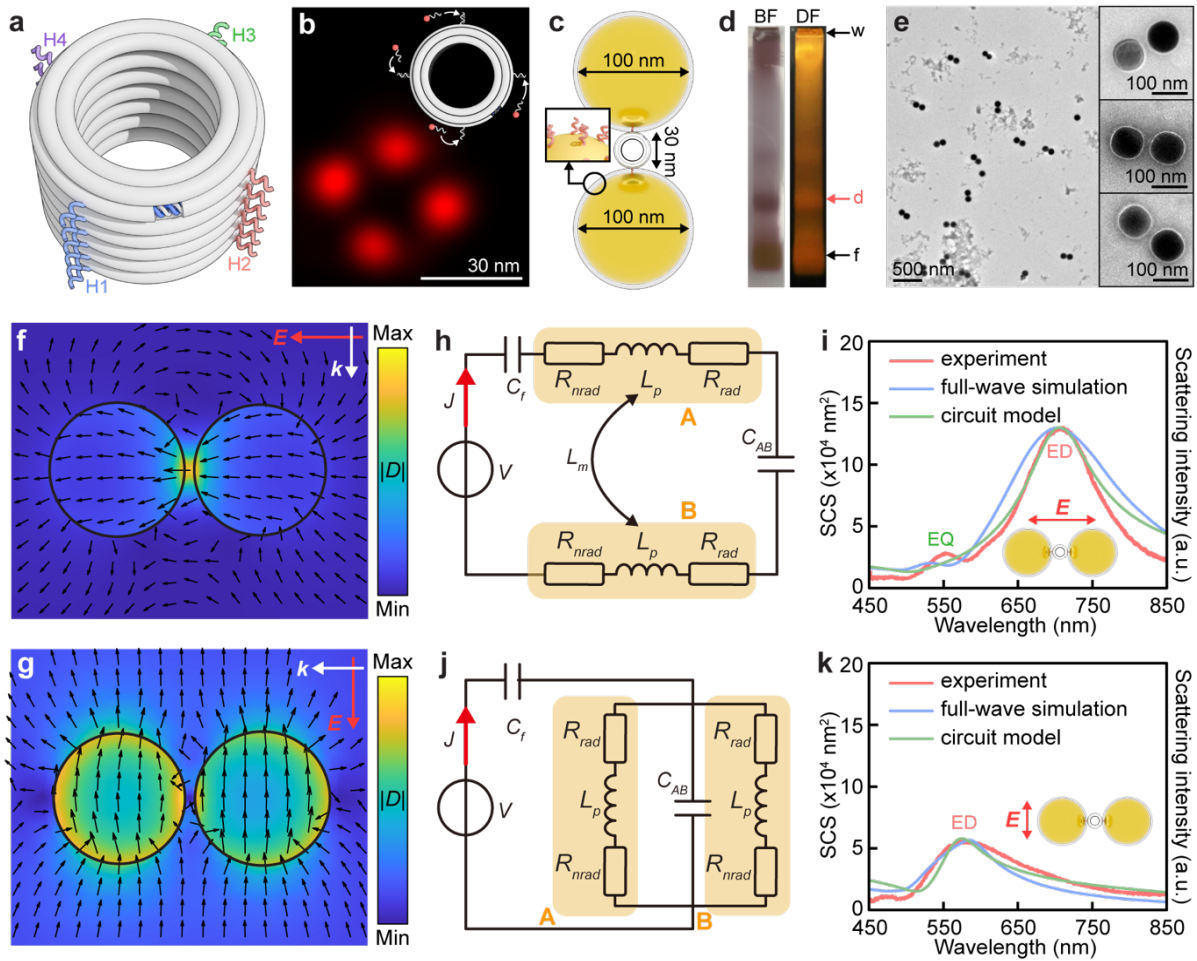

**Fig. 3 | Second order optical nanocircuit built with a symmetric Au NS dimer. a,** Schematic illustration of barrel DNA origami with four unique ssDNA handle sequences on the outer surface (H1-H4). **b,** Representative super-resolution optical image (DNA-PAINT) showing molecular addressability of ssDNA handles on barrel origami. **c,** Schematic illustration of the symmetric Au NS dimer configuration with a central barrel DNA origami. **d,** Image of agarose gel electrophoresis results for the symmetric dimer, imaged by bright-field (BF) and DF microscopy. Labels 'w', 'd', and 'f' labels indicate the gel well, symmetric dimer band, and free Au NSs, respectively. **e,** Wide-field TEM images of the purified symmetric dimer. The insets show the representative zoomed-in views. **f-i,** (**f**) Electric displacement field distribution at ED mode, (**h**) Thevenin equivalent nanocircuit diagram, and (**i**) scattering spectra of the symmetric dimer under horizontal illumination. $L_m$ indicates mutual inductance between $L_p$ of the two Au NSs in a dimer. **g-k,** (**g**) Electric displacement field distribution at ED mode, (**j**) Thevenin equivalent nanocircuit diagram, and (**k**) scattering spectra of the symmetric dimer under the vertical illumination.

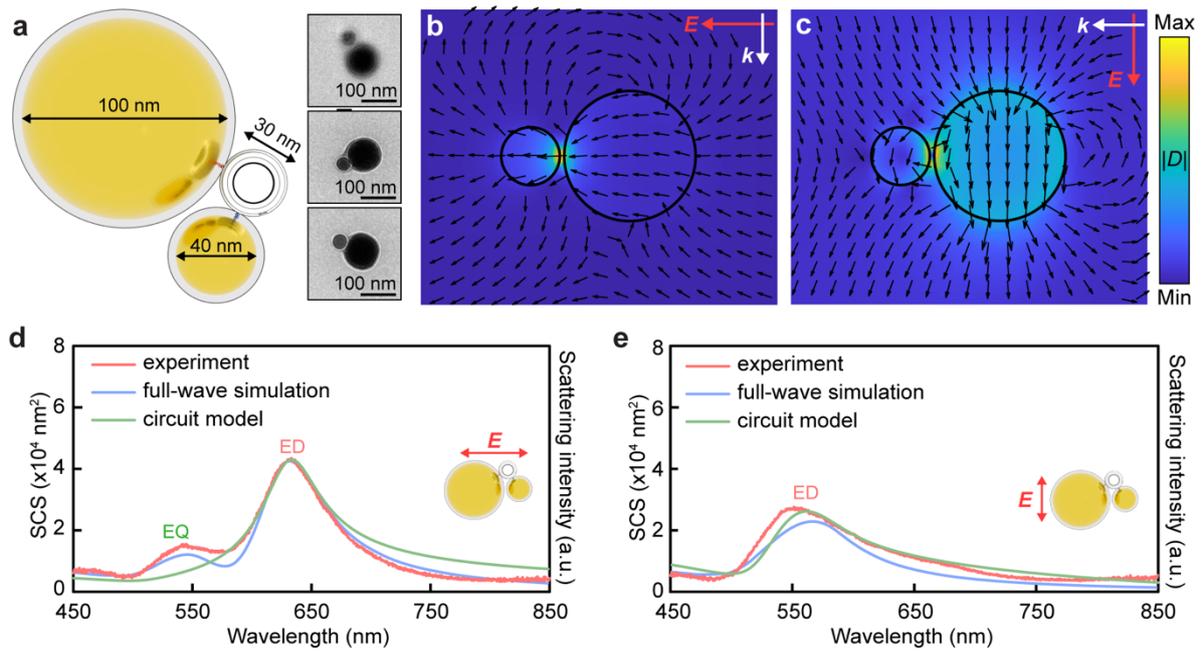

**Fig. 4 | Second order optical nanocircuit built with an asymmetric Au NS dimer. a,** (left panel) Schematic illustration of the asymmetric Au NS dimer configuration with barrel origami. (right panel) Representative TEM images of the assembled asymmetric Au NS dimers. **b-c,** Electric displacement field distribution of the asymmetric dimer under (**b**) horizontal and (**c**) vertical illumination. **d-e,** Theory predications and experimental measurements of the scattering spectra of the asymmetric dimer for (**d**) horizontal and (**e**) vertical illumination.

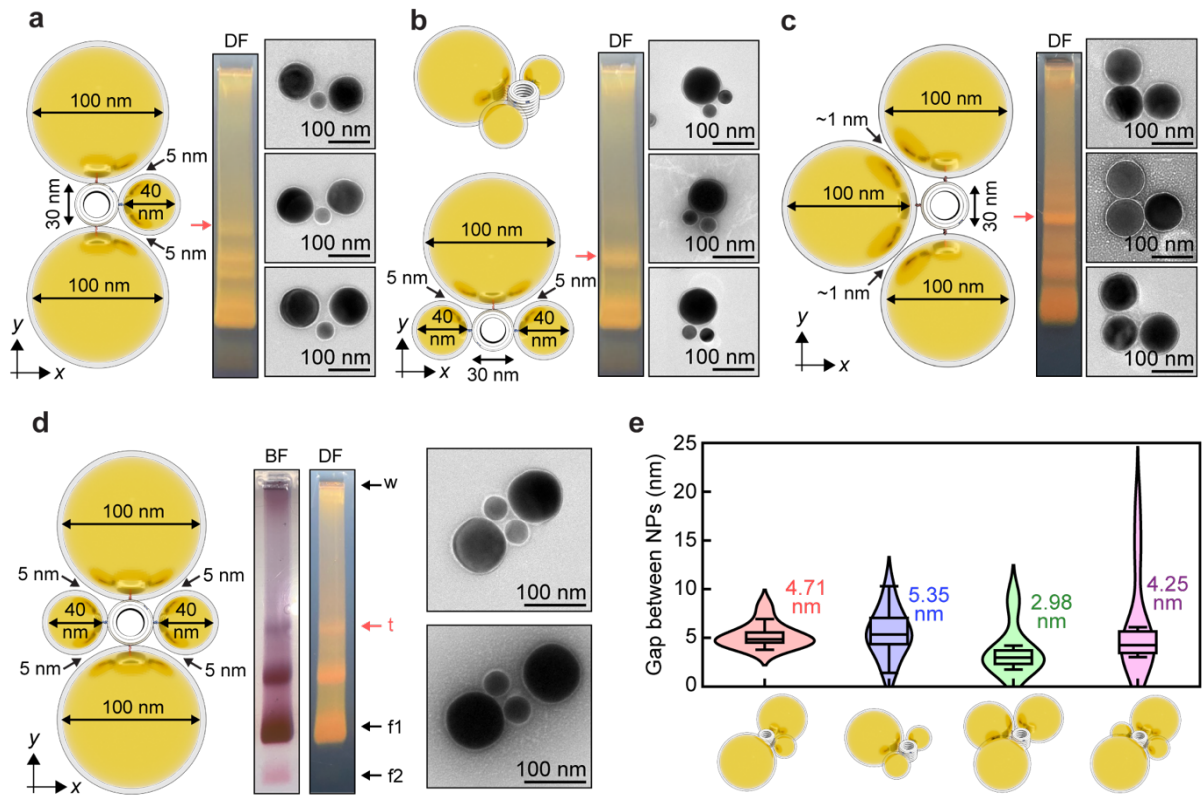

**Fig. 5 | Third and fourth order optical nanocircuit built with trimeric and tetrameric Au NS assemblies. a-d,** (left panel) Schematic illustration, (middle panel) agarose gel electrophoresis results with BF and DF imaging, and (right panel) representative TEM images of Au NS assemblies consist of (**a**) two 100 nm and one 40 nm Au NSs, (**b**) one 100 nm and two 40 nm Au NSs, (**c**) three 100 nm Au NSs, and (**d**) two 100 nm and two 40 nm Au NSs. Labels 't', 'f1' and 'f2' labels indicate target structure, 100 nm free Au NSs, and 40 nm free Au NSs, respectively. **e,** Violin and box plot for the gap distance between the assembled Au NSs ($N$ = 12 and 24 for trimers and tetramers, respectively). Numbers indicate the median value.

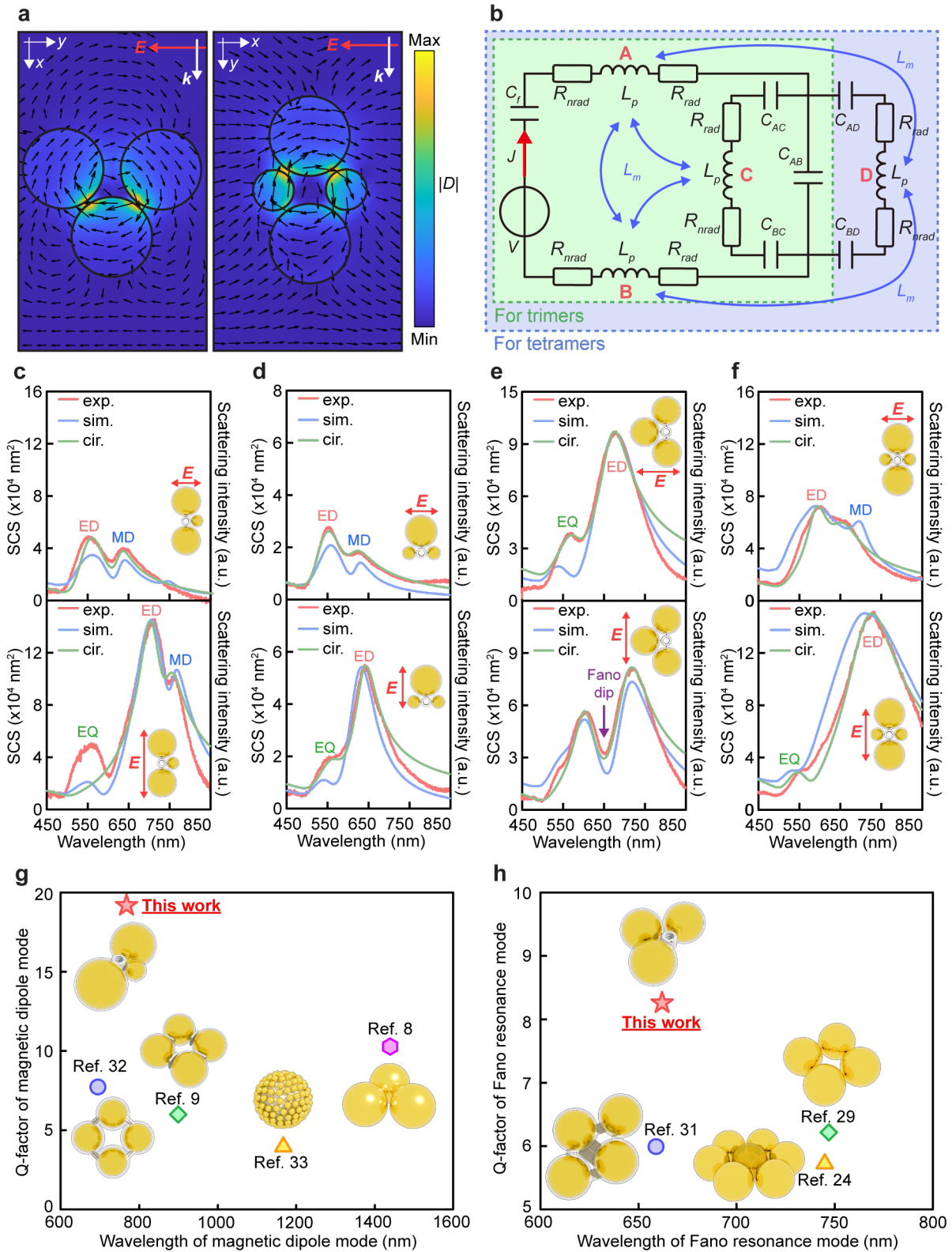

**Fig. 6 | Resonant properties of third and fourth order optical nanocircuit built with Au NS assemblies. a,** Electric displacement field distribution of (left panel) trimeric and (right panel) tetrameric Au NS assemblies, which induce a circulating electric displacement field. **b,** The third and fourth order Thevenin equivalent nanocircuit diagrams corresponding to trimeric and tetrameric Au NS assemblies, respectively. In this nanocircuit representation, green and

blue boxes correspond to trimeric and tetrameric circuits, while blue arrows highlight the mutual inductance ($L_m$). **c-f,** Scattering spectra of Au NS assemblies consist of (**c**) two 100 nm and one 40 nm Au NSs, (**d**) one 100 nm and two 40 nm Au NSs, (**e**) three 100 nm Au NSs, and (**f**) two 100 nm and two 40 nm Au NSs. exp., sim., and cir. correspond to the experiment, full-wave simulation, and circuit model analysis, respectively. **g-h,** Benchmark comparison of the third order optical nanocircuit with previously reported counterparts, focusing on the $Q$-factors of (**g**) MD and (**h**) Fano resonance modes (*8, 9, 24, 29, 31–33*).

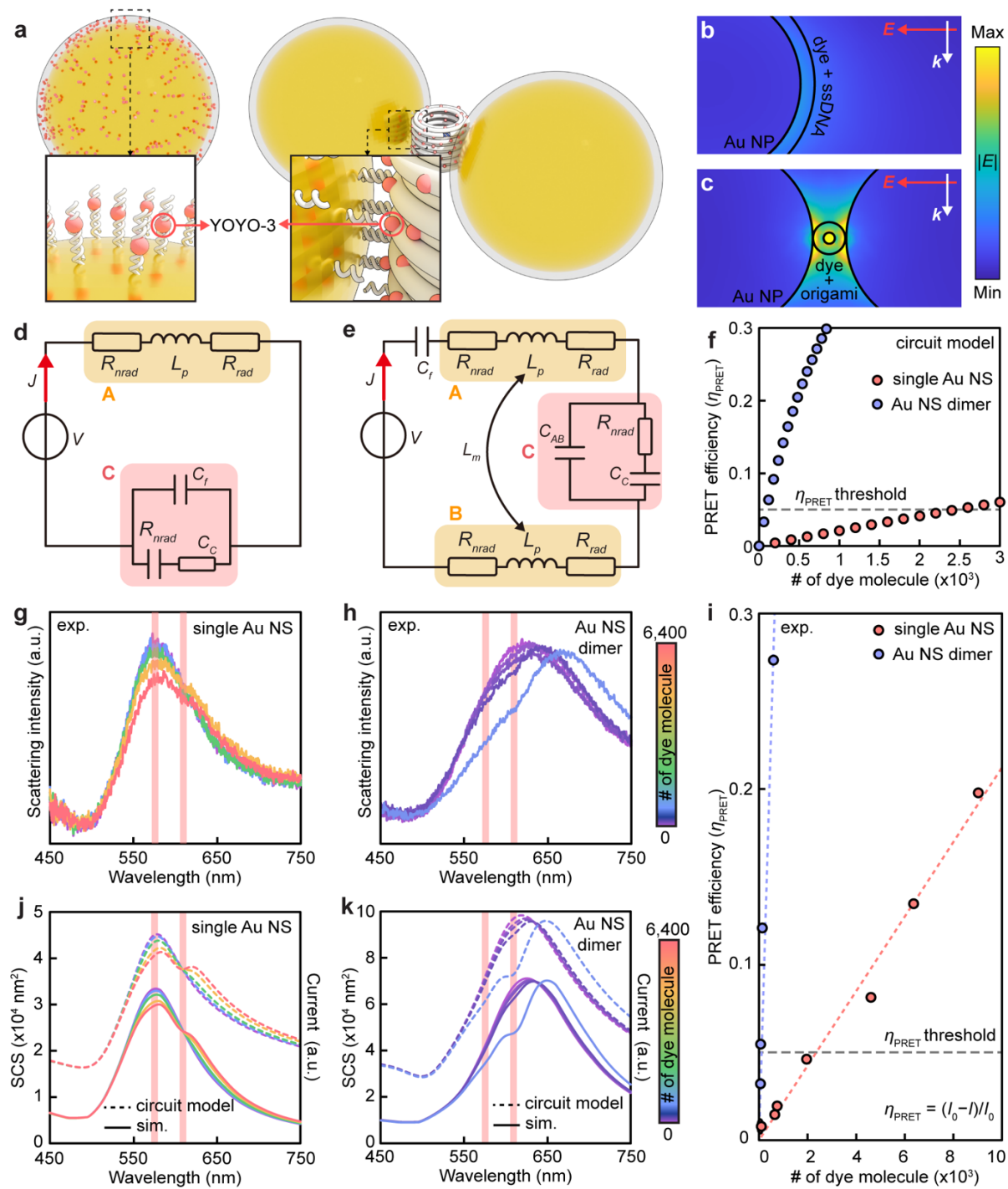

**Fig. 7 | Molecularly addressable optical nanocircuit for boosting plasmon resonance energy transfer (PRET) behavior. a,** Schematic representation of dye loaded Au NS (left panel) monomer and (right panel) dimer, which correspond to monomeric and dimeric optical nanocircuits. The red spheres indicate YOYO-3 dye molecules. **b-c,** Spatial distribution of electric field at ED modes of Au NS (**b**) monomer and (**c**) dimer. **d-e,** Schematic representation of the Thevenin equivalent optical nanocircuit diagram for dye-loaded Au NS (**d**) monomer and (**e**) dimer. **f,** PRET efficiency predicted from the nanocircuit models in (d) and (e). **g-h,** DF scattering spectra obtained from the dye loaded Au NP (**g**) monomer and (**h**) dimer under varying dye loading amounts. Pink vertical lines indicate the absorption peak of YOYO-3 dye

molecules. **i,** PRET efficiency calculated from the DF scattering spectra in (g) and (h). **j-k,** Numerically calculated scattering cross-section and electric currents predicted by optical nanocircuit models for the dye loaded Au NP (**j**) monomer and (**k**) dimer under varying dye loading amounts.

*Supplementary Information*

# 3D DNA Origami-Enabled Molecularly Addressable Optical Nanocircuit

*Jaewon Lee*[1,+], *Hayun Ahn*[1,+], *Kyung Hun Rho*[1], *Shelley F. J. Wickham*[2,*], *William M. Shih*[3,4*], and *Seungwoo Lee*[1,3,5,6,*]

[1]KU-KIST Graduate School of Converging Science and Technology, Korea University, Seoul 02841, Republic of Korea

[2]School of Chemistry, School of Physics, The University of Sydney Nano Institute, The University of Sydney, Sydney, NSW 2006, Australia

[3]Wyss Institute for Biologically Inspired Engineering, Harvard University, Boston, Massachusetts 02115, United States

[4]Department of Cancer Biology, Dana Farber Cancer Institute; Department of Biological Chemistry and Molecular Pharmacology, Harvard Medical School, Boston, Massachusetts 02215, United States

[5]Department of Integrative Energy Engineering (College of Engineering) and Department of Biomicrosystem Technology, Korea University, Seoul 02841, Republic of Korea

[6]Center for Opto-Electronic Materials and Devices, Post-Silicon Semiconductor Institute, Korea Institute of Science and Technology (KIST), Seoul 02792, Republic of Korea

*Email: shelley.wickham@sydney.edu.au

*Email: william.shih@wyss.harvard.edu

*Email: seungwoo@korea.ac.kr

+Equally contributed to this work

# Contents:

1. Barrel-shaped DNA origami

2. Coarse-grained molecular dynamics calculation of DNA origami

3. Estimation of YOYO-3 loading on DNA conjugated gold nanosphere (DNA-Au NS) and barrel DNA origami

4. Analytical analysis of optical nanocircuit model

5. Numerical calculation

6. Synthesis of 40-100nm Au NSs

7. Absorption and scattering cross-section (ACS and SCS) of 30-100 nm Au NSs

8. DNA-PAINT analysis of barrel DNA origami

9. Analysis of dark-field scattering spectra

# I. Barrel-shaped DNA origami

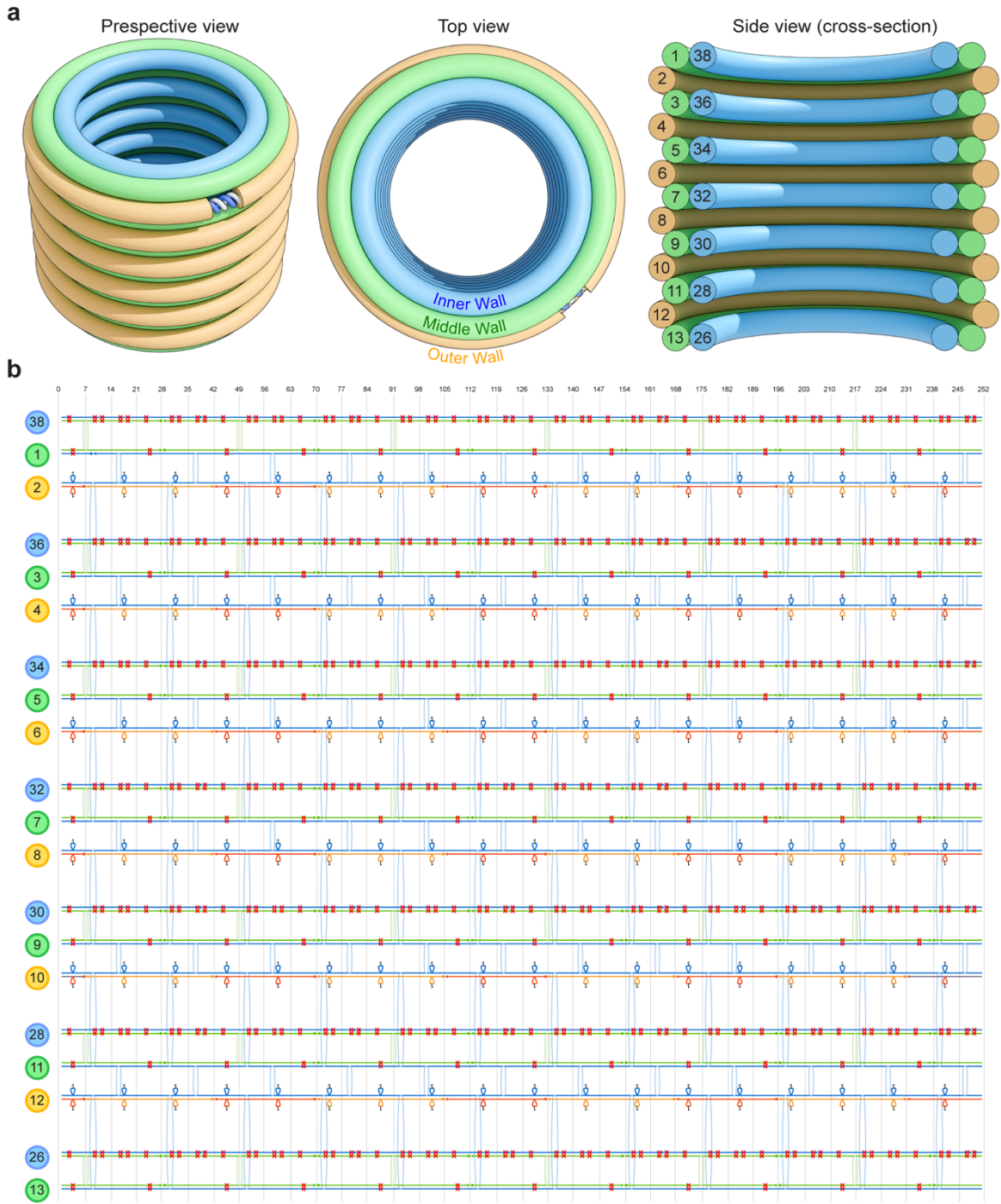

**Fig. S1 | Design and schematic of the barrel DNA origami structure. a,** Schematics of the DNA origami barrel shown from a perspective, top, and side view. The structure consists of three concentric walls of double-stranded DNA (dsDNA) helices, shown as cylinders. The inner, middle, and outer walls are colored blue, green, and orange, respectively. **b,** The caDNAno design scaffold and staple routing map for the barrel structure.

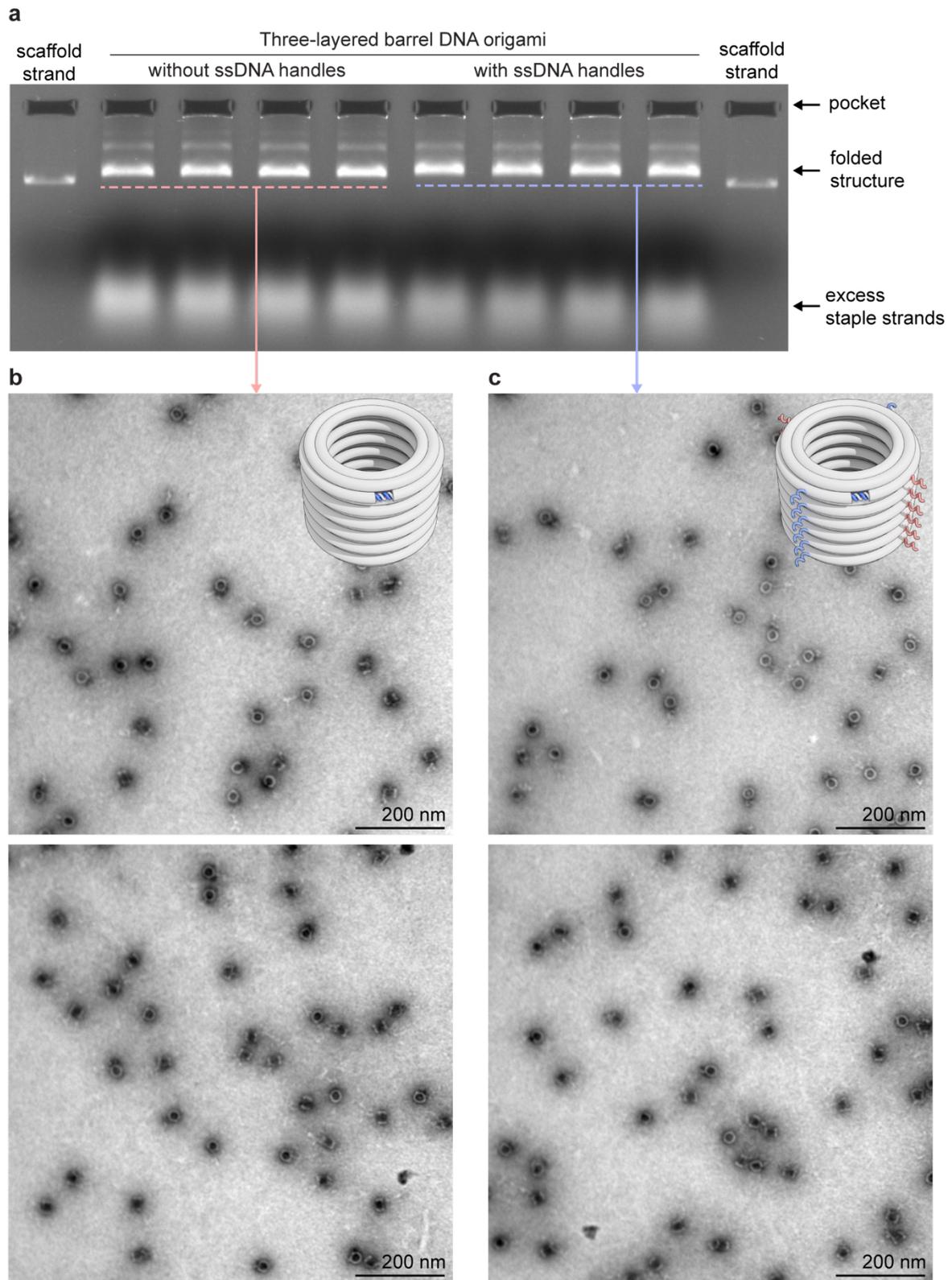

**Fig. S2 | Scalable assembly of barrel DNA origami. a,** Agarose gel electrophoresis analysis of the folded DNA origami. The gel confirms the successful formation of barrels without single-stranded DNA (ssDNA) overhangs and barrels functionalized with ssDNA handles. **b-c,** Representative low-magnification transmission electron microscopy (TEM) images of the assembled barrel DNA origami (b) without and (c) with ssDNA handles.

## 2. Coarse-grained molecular dynamics calculation of DNA origami

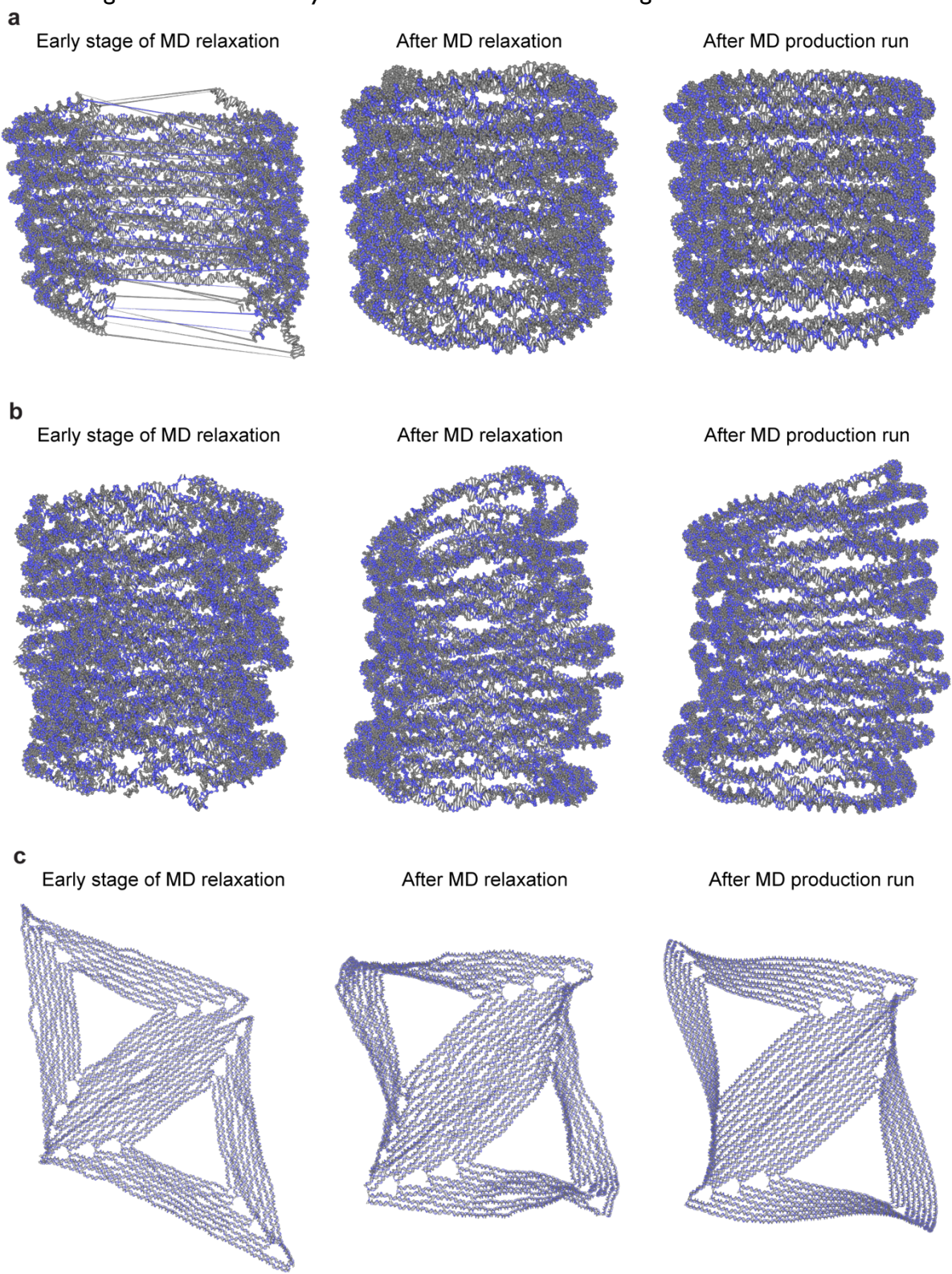

**Fig. S3 | Coarse-grained molecular dynamics (MD) simulation process of DNA origami structures. a-c,** Representative snapshots from MD simulations using the coarse-grained oxDNA model. The panels show the configurations of three different DNA nanostructures: (a) a three-layered barrel, (b) a two-layered barrel, and (c) a 2D rectangular DNA origami sheet.

In these coarse-grained DNA models, the scaffold strand is colored blue and the staple strands are grey.

Coarse-grained molecular dynamics (MD) simulations were performed using the oxDNA simulation package and followed established standard protocols (*1*, *2*). The simulation workflow consists of several stages. First, initial structural configurations of the DNA origami were generated from scadnano design files. These initial models were relaxed using a Monte Carlo (MC) simulation. Next, the structures underwent a two-step MD process. An initial MD relaxation was performed with modified backbone forces to allow structural rearrangement (see the early stage of Fig. S3). During this stage, a mutual trap potential was applied between base pairs to prevent the premature melting of double-stranded DNA (dsDNA) helices. After this initial relaxation, a final production run was conducted using the standard oxDNA force field without any external constraints (see the after MD relaxation and production run stages of Fig. S3). Finally, the simulation trajectories from the production run were analyzed to calculate the root mean square fluctuation (RMSF) and determine the mean structural configuration using the built-in oxDNA analysis tools (*2*).

## 3. Estimation of YOYO-3 loading on DNA conjugated gold nanosphere (DNA-Au NS) and barrel DNA origami

To estimate and control the amount of YOYO-3 dye loaded onto the DNA nanostructures, we first experimentally determined the dye's binding affinity for our specific buffer conditions. The relationship between the dye concentration in solution and the amount bound to DNA can be described by the McGhee-von Hippel model (*3*):

$$\frac{r}{c_f} = K_a(1-nr)\left[\frac{1-nr}{1-(n-1)r}\right]^{n-1} \qquad (1)$$

Here, $r$ is the ratio of bound dye to DNA base pairs, $c_f$ is the concentration of free dye in solution, $K_a$ is the association constant, and $n$ is the binding site size in base pairs (bp). The ratio $r$ and the free dye concentration $c_f$ can be expressed in terms of the total dye concentration ($c_T$), the bound dye concentration ($c_b$), and the total DNA base pair concentration ($c_{DNA}$):

$$c_b = c_T - c_f \qquad (2)$$

$$r = \frac{c_b}{c_{DNA}} = \frac{c_T - c_f}{c_{DNA}} \qquad (3)$$

$$\frac{c_T - c_f}{c_f c_{DNA}} = K_a\left(1 - n\frac{c_T - c_f}{c_{DNA}}\right)\left[\frac{1 - n\frac{c_T - c_f}{c_{DNA}}}{1 - (n-1)\frac{c_T - c_f}{c_{DNA}}}\right]^{n-1} \qquad (4)$$

Given a known binding site of $n=3.22$ bp/dye (*3*), the $K_a$ was the only unknown parameter. Since $K_a$ is highly sensitive to buffer conditions, we determined it experimentally.

To do this, we prepared solutions containing a fixed concentration of short (50 bp) double-stranded DNA (dsDNA) and titrated them with varying total concentrations of YOYO-3 ($c_T$). The $c_b$ was measured using absorption spectroscopy. Free and bound YOYO-3 have distinct absorption spectra, allowing for their quantification (*4*). The ($c_b$) was determined by first deconvoluting each spectrum with a two-component Lorentzian fit. This resolved the overlapping spectral contributions from free and bound dye. The center wavelength of each fitted peak was then used to calculate the relative abundance of the two species and thereby determine $c_b$.

The experimentally measured values for $c_b$ and $c_f$ at each titration point were then fitted to the McGhee-von Hippel model (Equation 1) to extract the association constant. For our experimental buffer, we determined the value to be $K_a=3.18\times10^4$ M$^{-1}$. This constant was then used to precisely estimate the YOYO-3 loading on the dsDNA-conjugated gold nanospheres (Au NSs) and the barrel DNA origami in our main experiments.

## 4. Analytical analysis of optical nanocircuit model

To analytically analyze the scattering and resonant behavior of various Au NS clusters and the PRET system, we used an optical nanocircuit model based on previously established frameworks (5, 6). The components were modeled based on the numerically simulated electric displacement field distribution depicted in Figure 2 of the main manuscript. A single Au NS was modeled as an inductor ($L_p$) coupled with a radiative resistor ($R_{rad}$) and a non-radiative resistor ($R_{nrad}$). The dye-loaded barrel DNA origami was modeled as a lossy capacitor, represented by a capacitor ($C$) coupled with a $R_{nrad}$.

Following previous work, the impedance of the Au NS and dye-loaded DNA origami can be represented as follows (5):

$$Z_{NP} = (-i\omega\varepsilon\pi R)^{-1} \qquad (5)$$

Here, $\omega$ is the angular frequency, $\varepsilon$ is the permittivity, and $R$ is the effective radius of the Au NS and DNA origami. The impedance for the displacement current of the fringe field can be represented as follows:

$$Z_{fringe} = (-i\omega 2\pi R\varepsilon_0)^{-1} \qquad (6)$$

$\varepsilon_0$ is the background permittivity. To model Au NS clusters, the circuit elements were integrated based on the cluster configuration and the polarization of the illuminating light. For a cluster's fringe field impedance, the effective radius $R$ was considered comparable to the cluster's dimension in the direction of the incident electric field. Additional capacitors ($C$) and mutual inductance ($l_m$) were introduced into the circuit to reflect the capacitive coupling between adjacent Au NSs. Because the measured scattered field can be related to the current delivered by the circuit's voltage source (6), the current flow in the final circuit was calculated to estimate the scattering and resonance behavior of the corresponding Au NS clusters and PRET systems. The nanocircuit parameters used for the lumped circuit elements and their assemblies are summarized in **Table S1-S5**.

**Table. S1 | Nanocircuit parameters used for dye-loaded DNA origami and Au NS.**

|  | $C_f$ (aF) | $r$ (nm) | $R_{rad}$ (Ω) |
|---|---|---|---|
| Au NS | 9.56 | 51 | 7 |
| Dye-loaded DNA origami | N/A | 15 | N/A |

**Table. S2 | Nanocircuit parameters used for dimeric nanocircuit. a,** Symmetric dimers composed of the identical 100 nm Au NSs. **b,** Asymmetric dimers composed of 100 nm and 40 nm Au NSs.

**a**

|  | $C_f$ (aF) | $r_A$ (nm) | $r_B$ (nm) | $R_{rad,A}$ (Ω) | $R_{rad,B}$ (Ω) | $l_m$ (fH) | $C_{AB}$ (aF) |
|---|---|---|---|---|---|---|---|
| x-polarized | 9.56 | 50.05 | 51.53 | 3.5 | 3.5 | 5 | 20 |

| | $C_f$ (aF) | $r_A$ (nm) | $r_B$ (nm) | $R_{rad,A}$ (Ω) | $R_{rad,B}$ (Ω) | $l_m$ (fH) | $C_{AB}$ (aF) |
|---|---|---|---|---|---|---|---|
| y-polarized | 7.82 | 50.05 | 51.53 | 1 | 1 | 0 | 10 |

**b**

| | $C_f$ (aF) | $r_A$ (nm) | $r_B$ (nm) | $R_{rad,A}$ (Ω) | $R_{rad,B}$ (Ω) | $l_m$ (fH) | $C_{AB}$ (aF) |
|---|---|---|---|---|---|---|---|
| x-polarized | 5.65 | 47.5 | 20 | 2 | 1 | 1 | 15 |
| y-polarized | 5.65 | 47.5 | 20 | 1 | 0.5 | 0 | 4 |

**Table. S3 | Nanocircuit parameters used for trimeric nanocircuit. a,** Trimers composed of two 100 nm Au NSs and one 40 nm Au NS with a mirror symmetry and heterogeneity **b,** Trimers composed of two 40 nm Au NS and one 100 nm Au NS with a mirror symmetry and heterogeneity. **c,** Trimers composed of three 100 nm Au NSs with a mirror symmetry without heterogeneity.

**a**

| | $C_f$ (aF) | $r_A$ (nm) | $r_B$ (nm) | $r_C$ (nm) | $R_{rad,A}$ (Ω) | $R_{rad,B}$ (Ω) | $R_{rad,C}$ (Ω) | $l_{m,AB}$ (fH) | $l_{m,BC}$ (fH) | $l_{m,AC}$ (fH) | $C_{AB}$ (aF) | $C_{BC}$ (aF) | $C_{AC}$ (aF) |
|---|---|---|---|---|---|---|---|---|---|---|---|---|---|
| x-polarized | 3.91 | 43.84 | 42.37 | 20 | 1 | 1 | 1 | 0 | 0.5 | 0 | 23 | 14 | 14 |
| y-polarized | 7.82 | 43.84 | 42.37 | 20 | 2 | 2 | 1 | 2.5 | 0.5 | 0.5 | 120 | 22 | 22 |

**b**

| | $C_f$ (aF) | $r_A$ (nm) | $r_B$ (nm) | $r_C$ (nm) | $R_{rad,A}$ (Ω) | $R_{rad,B}$ (Ω) | $R_{rad,C}$ (Ω) | $l_{m,AB}$ (fH) | $l_{m,BC}$ (fH) | $l_{m,AC}$ (fH) | $C_{AB}$ (aF) | $C_{BC}$ (aF) | $C_{AC}$ (aF) |
|---|---|---|---|---|---|---|---|---|---|---|---|---|---|
| x-polarized | 1.56 | 19.07 | 19.07 | 42.37 | 1 | 1 | 3 | 0 | 1 | 1 | 20 | 35 | 35 |
| y-polarized | 2.52 | 19.07 | 19.07 | 42.37 | 1 | 1 | 4 | 0 | 1.5 | 1.5 | 13 | 32 | 32 |

**c**

| | $C_f$ (aF) | $r_A$ (nm) | $r_B$ (nm) | $r_C$ (nm) | $R_{rad,A}$ (Ω) | $R_{rad,B}$ (Ω) | $R_{rad,C}$ (Ω) | $l_{m,AB}$ (fH) | $l_{m,BC}$ (fH) | $l_{m,AC}$ (fH) | $C_{AB}$ (aF) | $C_{BC}$ (aF) | $C_{AC}$ (aF) |
|---|---|---|---|---|---|---|---|---|---|---|---|---|---|
| x-polarized | 7.82 | 44.88 | 44.88 | 44.88 | 2 | 2 | 4 | 0 | 0.5 | 0.5 | 22 | 36 | 36 |
| y-polarized | 8.08 | 44.88 | 44.88 | 44.88 | 2.5 | 2.5 | 2 | 0.5 | 0.5 | 0.5 | 36 | 47 | 47 |

**Table. S4 | Nanocircuit parameters used for tetrameric nanocircuit.** Tetramers where two 100 nm Au NSs and two 40 nm Au NSs were also clustered into a mirror symmetry.

| | $C_f$ (aF) | $r_A$ (nm) | $r_B$ (nm) | $r_C$ (nm) | $r_D$ (nm) | $R_{rad,A}$ (Ω) | $R_{rad,B}$ (Ω) | $R_{rad,C}$ (Ω) | $R_{rad,D}$ (Ω) | $l_{m,AB}$ (fH) | $l_{m,BC}$ (fH) | $l_{m,AC}$ (fH) | $l_{m,AD}$ (fH) |
|---|---|---|---|---|---|---|---|---|---|---|---|---|---|
| x-polarized | 3.04 | 22.5 | 22.5 | 55 | 55 | 2.5 | 2.5 | 3 | 3 | 0 | 0.3 | 0.3 | 0.3 |
| | $l_{m,BD}$ (fH) | $C_{AB}$ (aF) | $C_{AC}$ (aF) | $C_{BC}$ (aF) | $C_{AD}$ (aF) | $C_{BD}$ (aF) | | | | | | | |
| | 0.3 | 55 | 80 | 80 | 80 | 80 | | | | | | | |

| | $C_f$ (aF) | $r_A$ (nm) | $r_B$ (nm) | $r_C$ (nm) | $r_D$ (nm) | $R_{rad,A}$ (Ω) | $R_{rad,B}$ (Ω) | $R_{rad,C}$ (Ω) | $R_{rad,D}$ (Ω) | $l_{m,AB}$ (fH) | $l_{m,BC}$ (fH) | $l_{m,AC}$ (fH) | $l_{m,AD}$ (fH) |
|---|---|---|---|---|---|---|---|---|---|---|---|---|---|
| y-polarized | 9.56 | 55 | 55 | 22.5 | 22.5 | 5 | 5 | 2 | 2 | 0 | 1.5 | 1.5 | 1.5 |
| | $l_{m,BD}$ (fH) | $C_{AB}$ (aF) | $C_{AC}$ (aF) | $C_{BC}$ (aF) | $C_{AD}$ (aF) | $C_{BD}$ (aF) | | | | | | | |
| | 1.5 | 16 | 16 | 16 | 16 | 16 | | | | | | | |

**Table. S5 | Nanocircuit parameters used for PRET nanocircuit. a,** dye loaded 100 nm Au NS. **b,** symmetric dimers composed of the identical 100 nm Au NSs.

**a**

| | $C_f$ (aF) | $r_A$ (nm) | $r_C$ (nm) | $R_{rad,A}$ (Ω) |
|---|---|---|---|---|
| x-polarized | 4.18 | 47.5 | 90 | 3.5 |

**b**

| | $C_f$ (aF) | $r_A$ (nm) | $r_B$ (nm) | $r_C$ (nm) | $R_{rad,A}$ (Ω) | $R_{rad,B}$ (Ω) | $l_m$ (fH) | $C_{AB}$ (aF) |
|---|---|---|---|---|---|---|---|---|
| x-polarized | 10.83 | 47.5 | 47.5 | 15 | 14 | 14 | 13.5 | 2.25 |

## 5. Numerical calculation

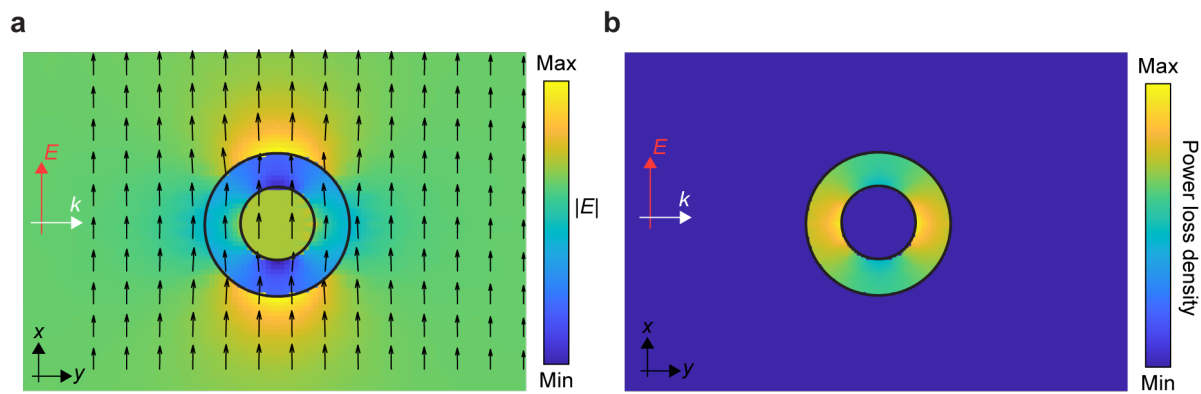

**Fig. S4 | Full-wave numerical calculation of dye-loaded DNA origami. a,** Electric field distribution of dye-loaded DNA origami at the wavelength of 610 nm. **b,** Power loss density distribution, revealing regions of optical absorption, of dye-loaded DNA origami at the wavelength of 610 nm.

The scattering and PRET behavior of the Au NS clusters were also analyzed using the finite element method (FEM) in a commercial software package (CST Microwave Studio, 2022). The dielectric function for Au was taken from the experimental data of Johnson and Christy (*7*). The specific geometry of each Au NS cluster, including NS sizes and inter-particle gap distances, was determined from corresponding TEM images. In all simulations, the clusters were illuminated by a plane wave at a 53° angle of incidence, and the linear polarization of the incident electric field was varied from 0° to 90°. The scattering cross section was then calculated from the power collected by a far-field monitor. The background medium was defined differently to account for each experimental condition. To model the dark-field scattering measurements on Formvar, the background was set as an effective medium with a dielectric constant of $\varepsilon = 1.56$, representing an average between air and the Formvar film. To model the PRET system within its microfluidic chamber, the background was set as an effective medium with a dielectric constant of $\varepsilon = 2.00$, representing an average between air and the glass substrate.

The dielectric properties of the DNA components were also specifically defined. The permittivity of the barrel-shaped DNA origami was determined to be 1.45 via the Clausius-Mossotti relation (*8*):

$$\varepsilon_{\text{DNA}} = \frac{3 + 2N\alpha_e}{3 - N\alpha_e} \varepsilon_h \tag{7}$$

Here, $\varepsilon_h$, $\varepsilon_{\text{DNA}}$, $N$, and $\alpha_e$ are the relative permittivity of the host medium, the relative permittivity of the DNA origami, the number density of nucleotides, and the electric polarizability of nucleotides, respectively. The ssDNA shell on the Au NSs was modeled with a thickness of 7.53 nm and a permittivity of 1.93, based on its optically quantified ssDNA loading density (*8*). For the PRET analysis specifically, the optical response of the dye molecules was included by modeling the dielectric function of the dye-loaded dsDNA shell and DNA origami with a Lorentzian oscillator model (*9*):

$$\varepsilon = \varepsilon_\infty + \frac{f_1 \omega_{0,1}^2}{\omega_{0,1}^2 - \omega^2 - i\gamma_{a,1}\omega} + \frac{f_2 \omega_{0,2}^2}{\omega_{0,2}^2 - \omega^2 - i\gamma_{a,2}\omega} \tag{8}$$

In this equation, $\varepsilon_\infty$ is the background dielectric constant, $\omega_0$ is the resonant angular frequency, and $\gamma_a$ is the linewidth of the dye molecule. The key parameter, $f$, is a dimensionless oscillator strength defined as:

$$f_i = \frac{N}{V}\mu^2 \frac{2}{3\varepsilon_0 \hbar \omega_{0,i}} \tag{9}$$

where $N$ is the number of dye molecules, $V$ is the volume of the DNA shell or origami, $\mu$ is the dipole moment of the dye, $\varepsilon_0$ is the vacuum permittivity, and $\hbar$ is the reduced Planck constant. The parameters for the YOYO-3 dye were determined from experimental measurements, using a two-oscillator model to capture its absorption profile: $\omega_{0,1}$ = 2.16 eV, $\omega_{0,2}$ = 2.03 eV; $\gamma_{a,1}$ = 0.20 eV, $\gamma_{a,2}$ = 0.12 eV; and $\mu$ = 8.5 D for both oscillators.

### 6. Synthesis of 40-100nm Au NSs

Au NSs with diameters ranging from 40 to 100 nm were synthesized following a modified, multi-step method of iterative reductive growth and oxidative dissolution (*8, 10, 11*). All chemical reagents were purchased from Sigma-Aldrich and used as received. The synthesis began with the preparation of a Au seed solution by injecting 125 µL of Gold (III) chloride trihydrate ($HAuCl_4$, 10 mM) into 5 mL of a hexadecyltrimethylammonium bromide (CTAB, 100 mM) solution under vigorous stirring. After 30 seconds, 300 µL of ice-cold sodium borohydride ($NaBH_4$, 10 mM) was rapidly injected. The solution was stirred for 1 minute and then incubated at 28°C for 30 minutes to decompose any residual $NaBH_4$.

This seed solution was then used to grow single-crystal Au nanorods (Au NRs). In a 200 mL solution of CTAB (100 mM), 10 mL of $HAuCl_4$ (10 mM) was injected, followed by the sequential addition of silver nitrate ($AgNO_3$, 1.8 mL, 10 mM), L-ascorbic acid (1.14 mL, 100 mM), and 240 µL of the seed solution, with 30-second intervals of mild stirring between each addition. After a minute of stirring, the solution was incubated at 28°C for 2 hours. Following growth, the ends of the Au NRs were etched out. To optimize this step, small batches of Au NRs (2 OD) in CTAB (50 mM) were incubated with varying concentrations of $HAuCl_4$ (from 50 to 100 µM) at 40°C for 2 hours. From these optimization experiments, the ideal $HAuCl_4$ concentration was typically found to be in the 70–80 µM range. This procedure was then repeated on a large scale using an optimized etchant concentration to produce the etched Au NRs (eAu NRs).

These eAu NRs were used as seeds to grow concave rhombic dodecahedra (CRDs). In 80 mL of a cetylpyridinium chloride (CPC, 20 mM) solution, 1.4 mL of $HAuCl_4$ (10 mM) and 18 mL of L-ascorbic acid (100 mM) were injected. Subsequently, 3 mL of the eAu NR seeds (1 OD), dispersed in CPC (10 mM), was added to the reaction solution and stirred for 15 minutes. To produce the 40 nm Au NSs, these CRDs (1 OD) were dispersed in CTAB (50 mM), mixed with $HAuCl_4$ to a final concentration of 60 µM, and incubated at 40°C for 2 hours.

The larger 100 nm Au NSs were prepared by using the 40 nm Au NSs as seeds in another growth and dissolution cycle. First, larger CRDs were grown by adding 4 mL of the 40 nm Au NSs (1 OD), dispersed in CPC (10 mM), to a solution containing CPC (80 mL, 20 mM), $HAuCl_4$ (1.4 mL, 10 mM), and L-ascorbic acid (18 mL, 100 mM), and stirring for 15 minutes. These larger CRDs (1 OD), dispersed in CTAB (50 mM), were then etched with $HAuCl_4$ (60 µM) at 40°C for 2 hours to produce the 100 nm Au NSs.

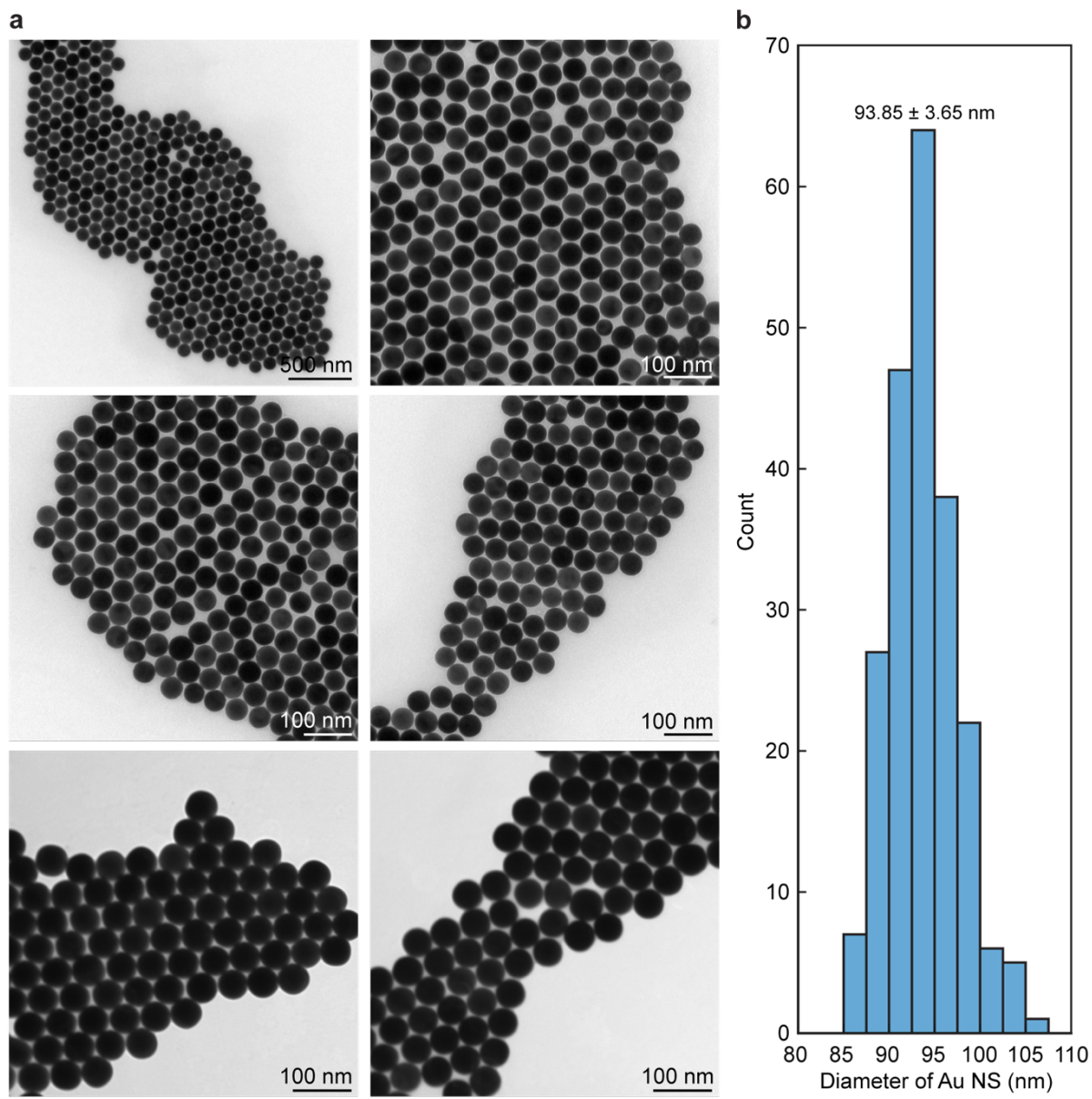

**Fig. S5 | Characterization of the synthesized 100 nm Au NSs. a,** Representative TEM images showing the uniform, highly spherical, and ultrasmooth features of Au NSs. **b,** The corresponding Au NS diameter distribution histogram, obtained from analysis of multiple TEM images.

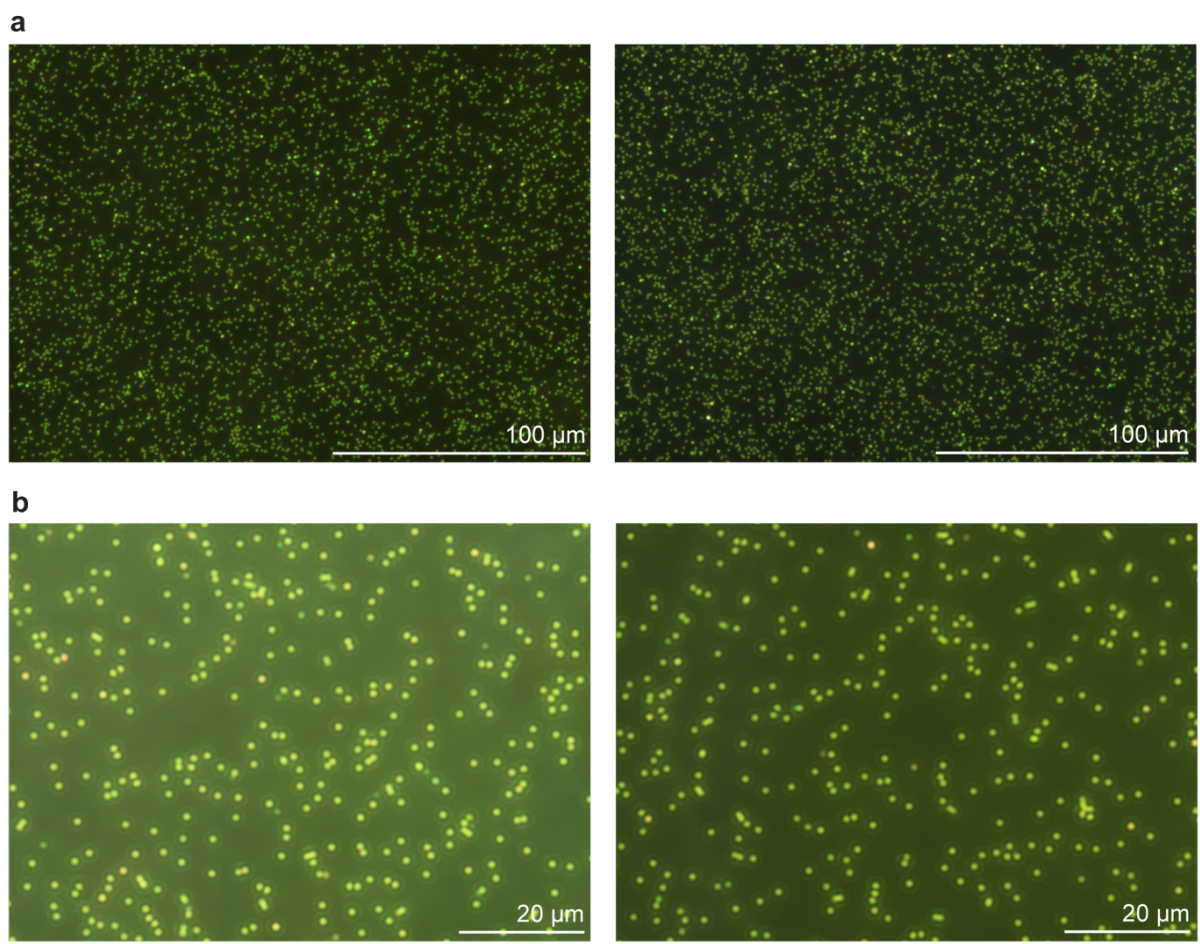

**Fig. S6 | Dark-field optical microscope images of 100 nm Au NSs. a-b,** Representative dark-field optical microscope images of Au NSs immobilized on the surface of a glass microfluidic chamber, viewed with **(a)** 10x and **(b)** 50x objective lens. The consistent green color and brightness of the scattered light from individual NSs indicate the high size and shape uniformity of the synthesized Au NSs.

## 7. Absorption and scattering cross-section (ACS and SCS) of 30-100 nm Au NSs

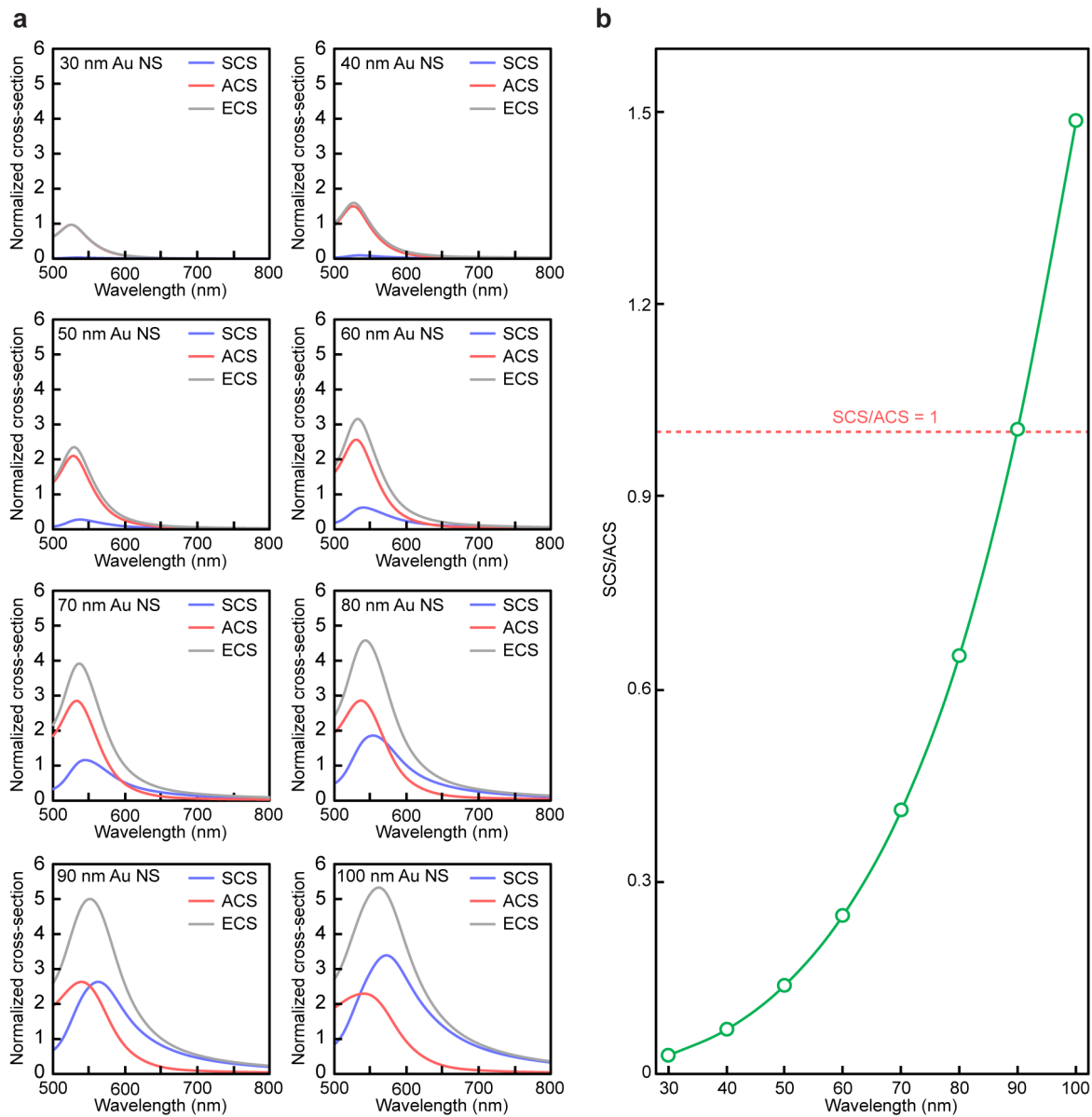

**Fig. S7 | Calculated cross-section of Au NSs as a function of diameter. a,** Numerically calculated absorption (ACS), scattering (SCS), and extinction (ECS) cross-sections for Au NSs with varying diameters. **b,** The corresponding ratio of SCS to ACS. The red dotted line highlights the crossover point where the scattering efficiency begins to dominate over absorption efficiency.

The absorption cross-section (ACS) and scattering cross-section (SCS) for Au NSs with diameters ranging from 30 to 100 nm were calculated using the FEM, as detailed in Section 5 of the supplementary information. To determine the ACS, the power loss density was integrated within the volume of the Au NS and then normalized by the intensity of the incident plane wave. For the SCS, the scattered electromagnetic field was first determined by subtracting the incident field from the total calculated field. The SCS was then found by integrating the

Poynting vector of this scattered field over a closed surface enclosing the nanosphere and normalizing the value by the incident plane wave intensity.

The calculated ACS, SCS, and their ratio (SCS/ACS) for different Au NS diameters are shown in **Figure S7**. For larger Au NSs (diameters exceeding 90 nm), the SCS becomes significantly greater than the ACS. This indicates that larger nanoparticles are more efficient at scattering energy than absorbing it. Therefore, plasmonic resonators assembled from these larger Au NSs can more effectively support complex, higher-order resonance modes (e.g., magnetic dipolar and Fano resonances) that rely on strong capacitive coupling between the Au NSs.

## 8. DNA-PAINT analysis of barrel DNA origami

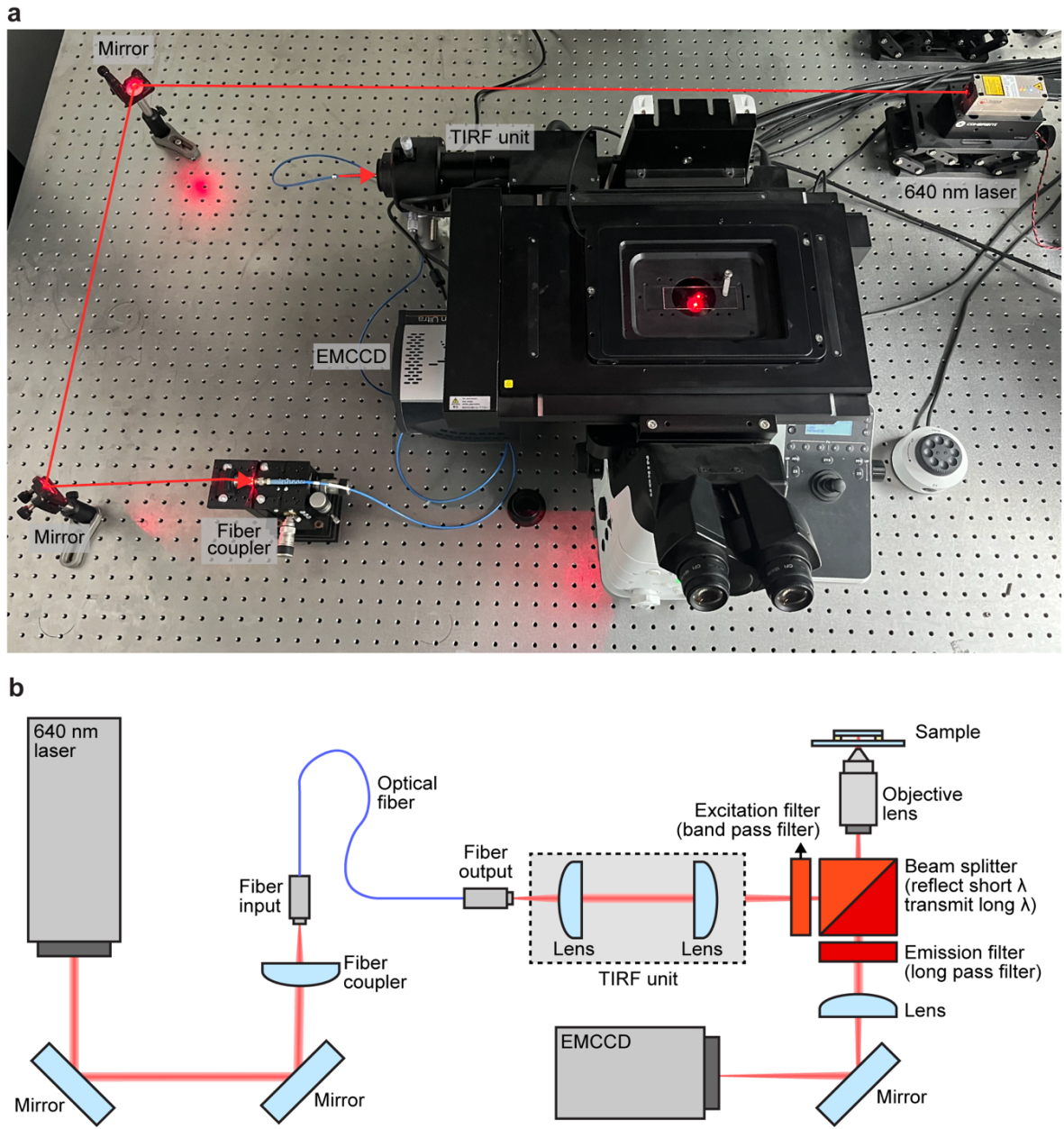

**Fig. S8 | Custom-built PAINT microscopy setup. a,** Photograph of the experimental setup showing the layout of the instrument. **b,** Corresponding schematic diagram illustrating the optical path and the arrangement of key components, such as the objective lens, filters, and detector.

DNA-PAINT imaging was performed on a custom-built microscopy setup, as illustrated in **Figure S8**. A 640 nm laser (TOPTICA, iBeam smart) was used as the excitation light source. The light was delivered through a polarization-maintaining single-mode optical fiber to ensure a high-quality Gaussian beam profile. The beam was directed through a TIRF illumination unit into a Nikon Ti2 series microscope. The light was then passed through a bandpass filter and beam splitter before being focused by a high numerical aperture objective lens (Nikon, Apo TIRF) to illuminate the sample. The resulting fluorescence emission was collected by the same objective lens, passed through an emission filter to block scattered excitation light, and imaged onto an EMCCD camera (Andor, iXon Ultra 888).

For the DNA-PAINT analysis of barrel DNA origami, the samples were specifically designed by incorporating 24 short docking strands on their outer surface (as shown in **Figure 3a** of the main manuscript) and 6 biotin molecules on their bottom surface. These structures were immobilized within a fluidic chamber via a biotin-streptavidin linkage. Briefly, the chamber surface was coated with BSA-biotin and then treated with streptavidin, following a previously reported protocol (*12*). Then, the biotin-tagged DNA origami was introduced and incubated to allow for immobilization on the surface. For imaging, a solution containing 5 nM of fluorescently-labeled imager strands, which are complementary to the docking strands, was introduced into the chamber. A total of 10,000 image frames were acquired using an exposure time of 250 ms per frame and without EM amplification. The final super-resolution images were reconstructed from the resulting localization data using the Picasso software package (*12*).

## 9. Analysis of dark field scattering spectra

### 9.1. Dark-field microspectroscopy set up

Scattering spectra were measured using a custom-built dark-field microspectroscopy system (*11*). Samples were illuminated with a broadband light source at an incident angle of 53°. The scattered light was collected through a 50x objective lens (NA 0.8) and directed to an imaging spectrometer (Princeton Instruments, IsoPlane) coupled with a CCD camera (Princeton Instruments, PIXIS-400B) for spectral analysis.

To investigate the resonant scattering behavior of the Au NS assemblies upon differently polarized light illumination, we performed polarization-resolved measurements. A linear polarizer was placed in the illumination path to set the polarization of the incident light. A second linear polarizer, acting as an analyzer, was placed in the detection path and aligned parallel to the first to selectively measure the co-polarized scattered light. Spectra were recorded by varying the incident polarization from 0° to 90°. For these measurements, 0° (horizontal) is defined as the orientation where the incident electric field is parallel to the long axis of the Au NS assembly, while 90° (vertical) corresponds to a perpendicular orientation.

### 9.2. Sample preparation and dark-field scattering measurement

To correlate the structure of Au NS clusters with their optical properties, assembled Au NS clusters were deposited onto a Formvar film supported by a 3-meshed TEM grid (*11*). This substrate was chosen because it enables TEM characterization prior to measuring its scattering spectrum. Furthermore, the optically thin Formvar film provides a high signal-to-noise ratio by minimizing background scattering. Other common substrates (e.g., glass and silicon wafers) were considered unsuitable for this correlative analysis. While glass offers a low scattering background, it is unsuitable for characterizing the target clusters with an electron microscope before the optical measurement. Silicon wafers, though compatible with both techniques, have a high refractive index that can distort the plasmonic resonances through mirror image of the induced dipole (*11*).

The scattering spectra measurement procedure began with a pre-screening step to ensure the reliability of the data acquisition. The TEM grid was first inspected using the dark-field optical microscope to select regions with low background noise, minimal contamination, and an intact Formvar film. Once a suitable region was identified, it was imaged using low-magnification TEM to create a map of well-isolated nanoclusters. Following this, intermediate-magnification TEM was used to determine the precise orientation of individual target clusters. High-magnification imaging was avoided to minimize potential electron beam-induced damage to the sample before optical characterization. Subsequently, the same TEM grid was transferred to the dark-field microscopy stage. Using the low-magnification TEM map as a guide, each structurally-characterized cluster was relocated. Finally, the polarization-dependent scattering spectrum of the single, pre-characterized cluster was measured. A representative example of this correlative imaging process is shown in **Figure S9**.

For characterization of the PRET system, a microfluidic chamber was constructed from two sandwiched soda-lime glass slides. One inner glass surface was treated with oxygen plasma to create a hydrophilic surface, which facilitated the immobilization of the dye-loaded Au NS monomers and dimers. This microfluidic device allowed for in-situ control over the sample environment during the scattering measurement. Once the nanostructures were immobilized, the buffer solution in the chamber was exchanged with solutions containing various concentrations of YOYO-3 dye. This process enabled precise control over the dye-loading density on the Au NS monomers and dimers while the scattering spectra were being measured.

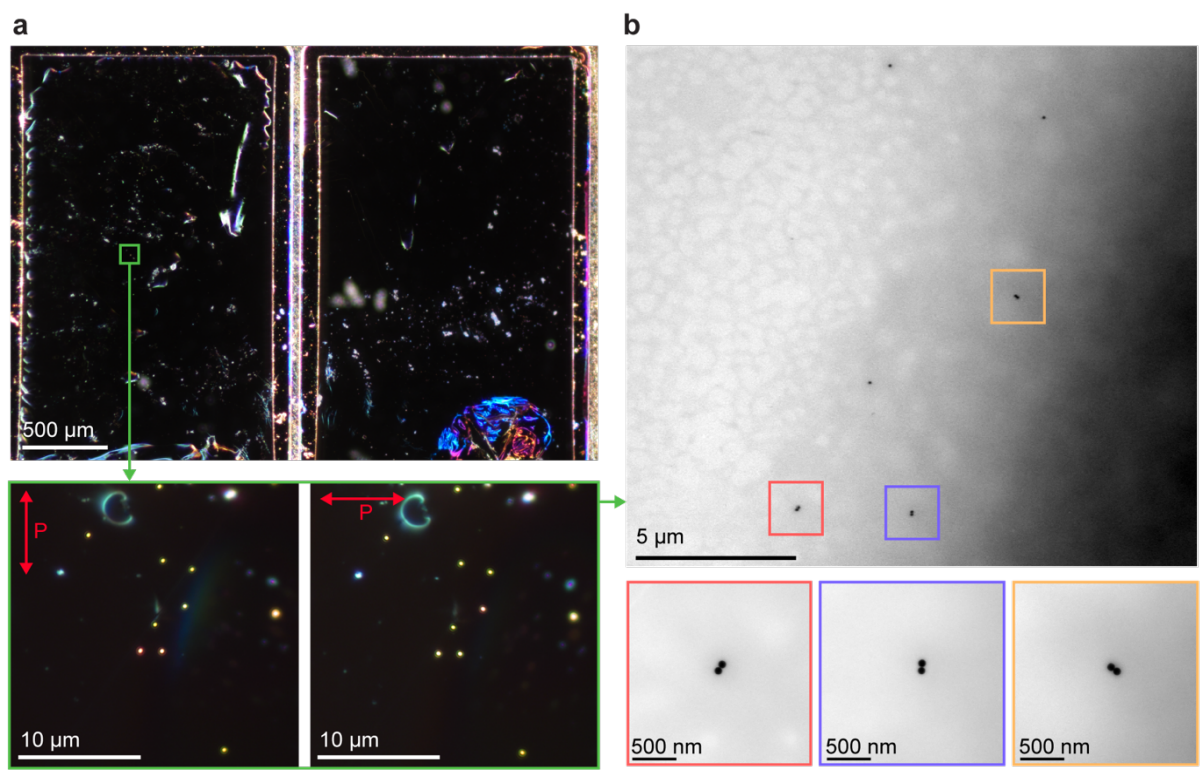

**Fig. S9 | Representative example of correlative imaging for dark-field scattering measurement. a,** Representative dark-field optical microscope images of Au NS dimers on a Formvar film supported by a 3-meshed TEM grid. The arrow labeled 'P' indicates the polarization of the incident light. **b,** Corresponding TEM images of the exact same region. The top panel shows a low-magnification TEM map. The bottom panel provides an intermediate-magnification view, which is used to determine the precise orientation of the target dimer before spectral measurement.

## 10. Preparation of DNA functionalized Au NSs

To functionalize the Au NSs with DNA, we utilized the stable pseudo-covalent bond formed between gold and a thiol group tagged at the end of a DNA oligo. The as-synthesized Au NSs, initially dispersed in CTAB (10 mM), were first washed to remove residual surfactant by concentrating and redispersing them in deionized (DI) water. The washed Au NSs were then dispersed in a buffer solution containing 0.05 wt% sodium dodecyl sulfate (SDS) and 10 mM Tris-HCl, along with the thiolated anti-handle ssDNA. This anti-handle DNA is complementary to the handle strands decorated on the surface of the barrel DNA origami. The concentration of the anti-handle ssDNA was maintained above 3 µM throughout the procedure. After a 1-hour incubation in a mixer, a salt-aging technique was employed to maximize the DNA loading density on the Au NS surface (*8*, *13*, *14*). This technique reduces the electrostatic repulsion between DNA strands. A high-salt buffer (2 M NaBr, 0.05 wt% SDS, 10 mM Tris-HCl) was gradually injected into the solution over 2 days until the final NaBr concentration reached 750 mM.

Successful DNA conjugation was confirmed by observing a 1-2 nm redshift in the extinction spectra of the DNA-conjugated Au NSs compared to the bare particles. The successful functionalization was further supported by a salt screening assay, which assesses the colloidal stability of the NSs in high-salt solutions. Finally, before their use in cluster assembly, the DNA-conjugated Au NSs were washed by cetnrifugating and redispersing them in buffer (0.05 wt% SDS, 10 mM Tris-HCl) to remove any excess anti-handle ssDNA. The thiolated anti-handle ssDNA was purchased from Bioneer, and all other chemical reagents were ordered from Sigma-Aldrich.

11. Electric quadrupolar mode of Au NS dimer

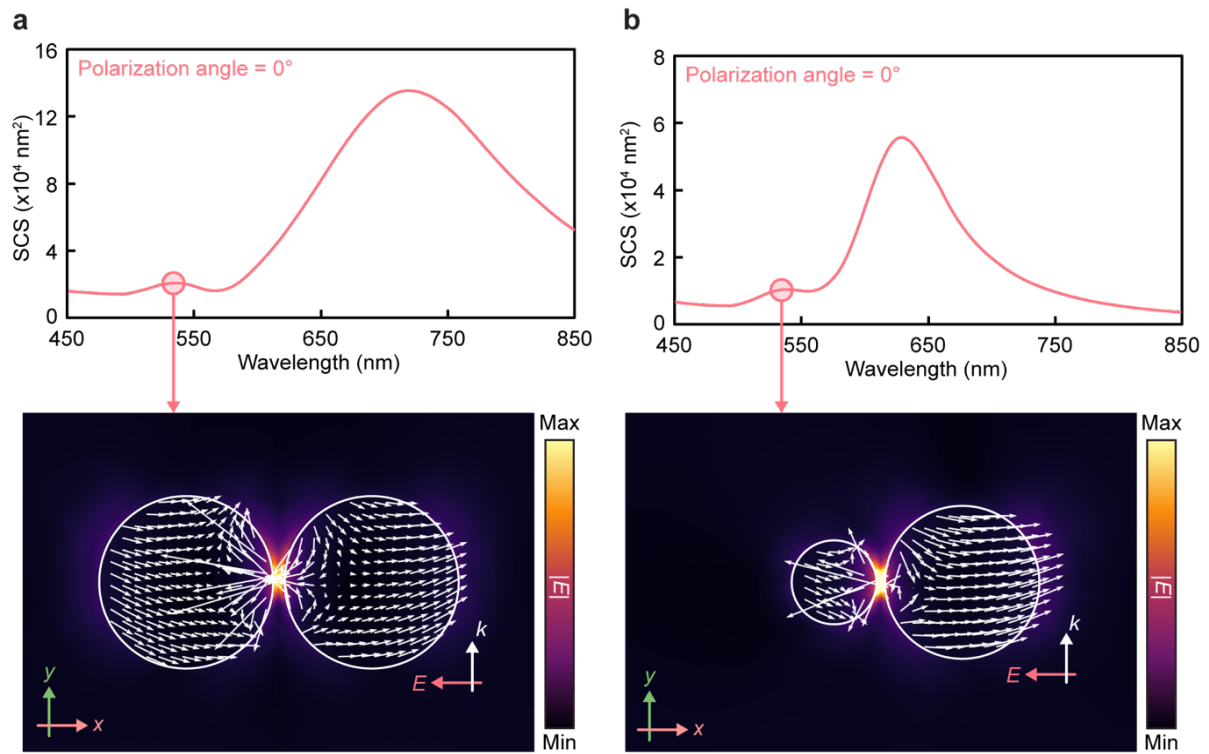

**Fig. S10 | Electric field distribution of the electric quadrupolar (EQ) mode in Au NS dimers. a-b,** (Top panel) numerically simulated SCS for (**a**) a symmetric and (**b**) an asymmetric Au NS dimer. (Bottom panel) Corresponding electric field distributions calculated at each dimer's respective EQ resonance wavelength. The white arrows indicate the internal electric field vectors within the nanoparticles. This complex, non-dipolar field pattern at the EQ mode cannot be captured by the fundamental dipolar approximation used in the lumped optical nanocircuit model.

## 12. *Q*-factor analysis of electric and magnetic resonance modes

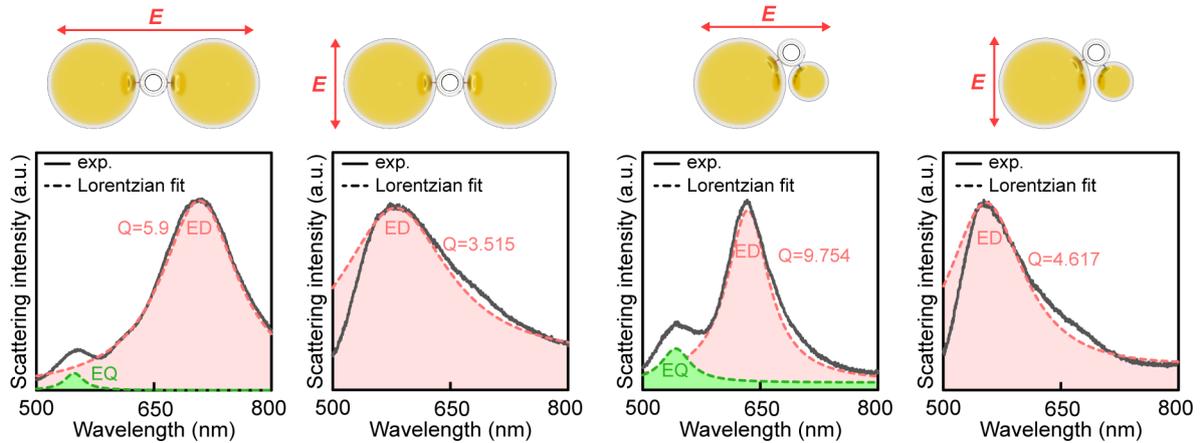

**Fig. S11 | Q-factor analysis of the electric dipole (ED) resonance of Au NS dimers.** (Top panel) Schematic illustration of the Au NS dimer under different incident polarization. (Bottom panel) Representative measured scattering spectrum (solid line) from an assembled dimer. The dashed line shows a Lorentzian fi t applied to the resonance modes, which is used to extract its Q-factor.

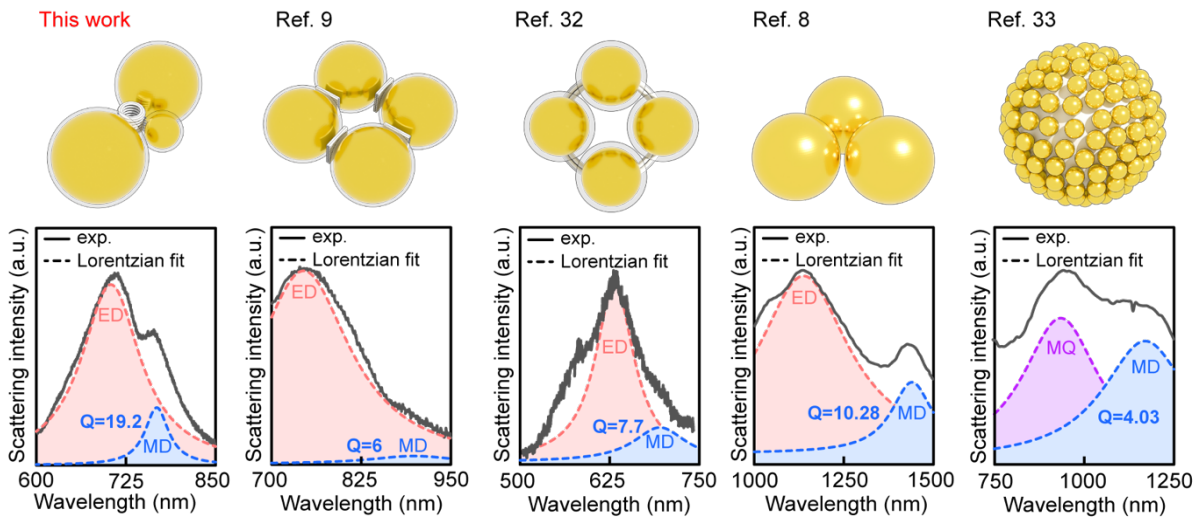

**Fig. S12 | Benchmarking the Q-factor of the magnetic dipole (MD) mode of a trimeric Au NS cluster.** (Top panel) Schematic illustrations of the trimeric Au NS cluster (this work) and other cluster geometries that exhibit MD mode. (Bottom panel) Comparison of scattering spectra and their corresponding Q-factors. The spectrum for the trimer was measured in this work, while the spectra for the benchmark clusters were obtained from published data referred in the main manuscript. To ensure a consistent comparison, a Lorentzian function (dashed lines) was fitted to the resonance modes of each spectrum to extract and compare their respective Q-factors.

## 13. *Q*-factor analysis of Fano resonance mode from symmetry-broken trimer

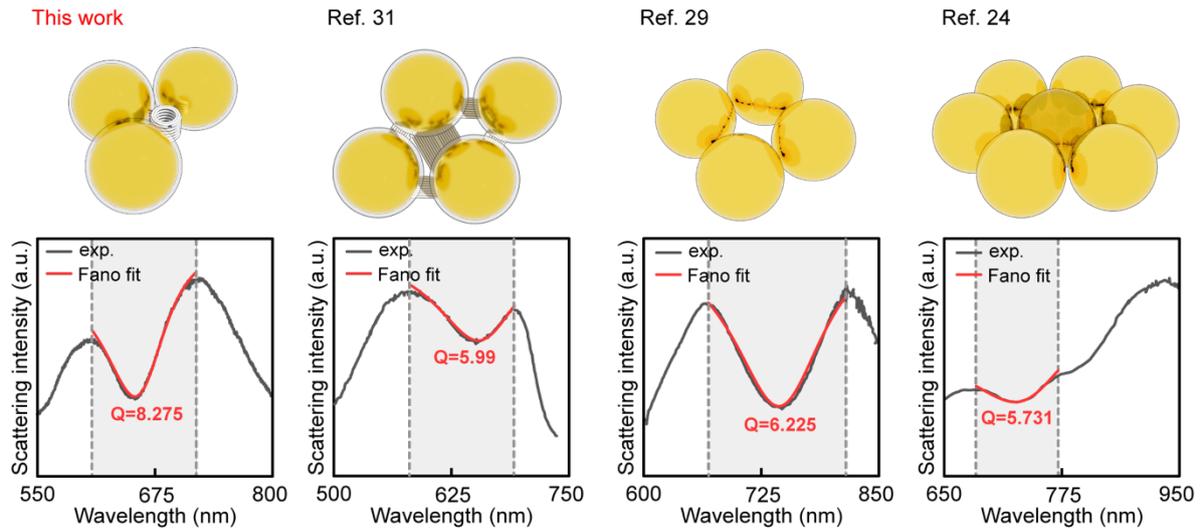

**Fig. S13 | Benchmarking the Q-factor of the Fano resonance mode of a symmetry-broken Au NS trimer.** (Top panel) Schematic illustrations of the symmetry-broken trimeric Au NS cluster (this work) and other cluster geometries that exhibit a Fano resonance mode. (Bottom panel) Comparison of scattering spectra and their corresponding Q-factors. The spectrum for the trimer was measured in this work, while the spectra for the benchmark clusters were obtained from published data referred in the main manuscript. To ensure a consistent comparison, a Fano function (red solid lines) was fitted to the resonance mode of each spectrum within the shaded grey region to extract and compare their respective quality factors.